\documentclass[amsmath, amssymb, aps, prb, longbibliography, superscriptaddress, twocolumn, floatfix, 10pt, nofootinbib]{revtex4-1}
\usepackage[utf8]{inputenc}
\usepackage{gensymb,graphicx,color}
\usepackage{hyperref}
\usepackage{multirow}
\usepackage{bm}
\usepackage[normalem]{ulem}
\usepackage[caption=false]{subfig}

\AtBeginDocument{\renewcommand{\natexlab}[1]{#1}}

\newcommand{\refcite}[1]{Ref.\,\onlinecite{#1}}

\newcommand{\eqnref}[1]{Eq.\,(\ref{#1})}

\newcommand{\figref}[1]{Fig.\,\ref{#1}}
\newcommand{\sfigref}[2]{Fig.\,\hyperref[#1]{\ref{#1}(#2)}}
\newcommand{\tabref}[1]{Tab.\,\ref{#1}}

\newcommand{\appref}[1]{Appendix\,\ref{#1}}
\newcommand{\newsec}[1]{{\bf \emph{#1}}---}

\definecolor{pink}{RGB}{200,0,200}

\hypersetup{
  colorlinks = true,
  urlcolor = blue,
  pdfauthor = {Kevin Slagle and Liang Fu},
  pdftitle = {Slagle and Fu - 2020 - Charge Transfer Excitations, Pair Density Waves, and Superconductivity in Moire Materials}
}

\begin{document}

\title{Charge Transfer Excitations, Pair Density Waves, and Superconductivity in \\ Moir\'e Materials}
\date{\today}

\author{Kevin Slagle}
\affiliation{Department of Physics and Institute for Quantum Information and Matter, California Institute of Technology, Pasadena, California 91125, USA}
\affiliation{Walter Burke Institute for Theoretical Physics,
California Institute of Technology, Pasadena, California 91125, USA}
\author{Liang Fu}
\affiliation{Department of Physics, Massachusetts Institute of Technology, Cambridge, MA 02139 USA}

\begin{abstract}
Transition metal dichalcogenide (TMD) bilayers are a new class of tunable moir\'e systems attracting interest as quantum simulators of strongly-interacting electrons in two dimensions.
In particular, recent theory predicts that the correlated insulator observed in WSe$_2$/WS$_2$ at half filling is a charge-transfer insulator similar to cuprates
  and, upon further hole doping, exhibits a transfer of charge from anion-like to cation-like orbitals at different locations in the moir\'e unit cell.
In this work, we demonstrate that in this doped charge-transfer insulator, tightly-bound charge-$2e$ excitations can form to lower the total electrostatic repulsion.
This composite excitation, which we dub a \emph{trimer}, consists of a pair of holes bound to a charge-transfer exciton.
When the bandwidth of doped holes is small, trimers crystallize into insulating pair density waves at a sequence of commensurate doping levels.
When the bandwidth becomes comparable to the pair binding energy, itinerant holes and charge-$2e$ trimers interact resonantly, leading to unconventional superconductivity similar to superfluidity in an ultracold Fermi gas near Feshbach resonance.
Our theory is broadly applicable to strongly-interacting charge-transfer insulators,
  such as WSe$_2$/WS$_2$ or TMD homobilayers under an applied electric field.
\end{abstract}

\maketitle

Moir\'e superlattices \cite{MacDonaldGraphene,BalentsMoire,CastroTwist} can be viewed as magnified crystals whose unit cell is of nanometer instead of angstrom size.
Correspondingly, the relevant electronic phenomena in moir\'e superlattices is governed by the coarse-grained moir\'e potential and the extended Coulomb repulsion, with characteristic energy scale on the order of meV instead of eV. Thanks to the increase of length scale and the reduction of energy scale, moir\'e systems feature remarkable tunability through the control of twist angle and displacement field. A variety of moir\'e materials have emerged as exciting venues for studying and designing correlated electron phenomena with an unprecedented level of control.
  \cite{Pablo2018,PabloInsulator,Dmitri2019,Wang2018,NadjPerge2019,Zhang2019,CarrPoWannier,PoModels,FuWannier,KangTBG,TarnopolskyMagic,XuWire}

Recently, moir\'e superlattices of transition metal dichalcogenides (TMD) \cite{TMDRev} have attracted great interest as quantum simulators of strongly-correlated electron systems in two dimensions. \cite{DeanTMD,DeanHallTMD,DeanExciton,FengExciton,Imamoglu2019,Park2020}
By varying the twist angle, the relative strength of the bandwidth and electron interaction can be tuned, and a rich quantum phase diagram can potentially be realized. \cite{BiFuExcitonicTMD,MacDonaldTMD,MacDonaldTopoTMD,TMDRealization}
Encouragingly, transport and optical experiments are starting to observe correlated insulating states in the TMD heterobilayer WSe$_2$/WS$_2$ with $n \leq 1$ holes per moir\'e unit cell. \cite{FaiTMD, BerkeleyTMD,ShanTMD}
In particular, the insulating state at $n=1$ is theoretically identified  as a charge-transfer insulator with a cation and an anion at different locations in the moir\'e unit cell,
  corresponding to localized Wannier orbitals at the primary and secondary energy minimum of the moir\'e potential, respectively \cite{FuChargeTransfer,Efoot:graphene}.
While a charge-transfer insulator is similar to a Mott insulator in terms of ground state properties, the key difference is that upon doping a charge-transfer insulator to $n>1$, the additional charges fill a higher-energy orbital in order to avoid double occupancy \cite{chargeTransferInsulator}. A famous example of charge-transfer insulators is undoped cuprates \cite{EmeryHighTc,ZhangRiceCuO}, for which a link between charge-transfer physics and high-temperature superconductivity upon doping has long been proposed and studied \cite{BarYamTwoComponent,VarmaChargeTransfer,KotliarChargeTransfer}.

In this work, we present a microscopic theory of charge pairing by Coulomb repulsion in TMD heterobilayers under a range of fillings $n>1$.
This counter-intuitive phenomenon occurs when the charge-transfer gap at $n=1$ is small, so that two doped charges can lower their energy by polarizing their surroundings to form a tightly-bound charge-$2e$ ``trimer'' that consists of three holes on adjacent cations surrounding an electron on an anion.
We show that the trimer costs less energy than two spatially separated holes for realistic forms of electron-electron interaction.
When the single-particle bandwidth is small, we predict the formation of periodic density waves of trimers at certain doping levels $n=1+ p/q>1$ ($p,q$ are integers), whose periodicity is commensurate with the moir\'e lattice.
As the bandwidth of holes increases and becomes comparable with the binding energy of trimers,
  holes and trimers coexist and interact resonantly to form a strong-coupling superconductor,
  similar to a strongly-interacting superfluid in a Fermi gas near Feshbach resonance \cite{GurarieRadzihovskyFeshbach,BECBCS}.
Our theory of pair density waves and superconductivity in TMD heterobilayers is asymptotically exact in a certain regime of strong interaction and small doping.

We start by describing the single-particle electronic structure of TMD heterobilayers (e.g.,  WSe$_2$/WS$_2$).
Here, the topmost valence band of WSe$_2$ is reconstructed into a set of moir\'e bands by the periodic moir\'e potential resulting from the $4\%$ lattice mismatch with WS$_2$.
Importantly, the moir\'e potential has two inequivalent local minima located at the AA and AB stacking regions, giving rise to two sets of localized Wannier orbitals.
The AA (AB) orbital has a lower (higher) on-site energy for holes and can be regarded as anion (cation) like.

Thus, the low-energy physics of a TMD heterobilayer can be faithfully mapped onto a two-dimensional diatomic crystal with one cation and one anion per unit cell. The effective Hamiltonian takes the form of an extended Hubbard model on the honeycomb lattice \cite{FuChargeTransfer}:
\begin{equation}
\begin{split}
  H &= H_K + H_{0} \hspace{1.8cm}
  H_K = \sum_{ij,s=\uparrow\downarrow} t_{ij} \, c^{\dagger}_{is}c_{js} \\
  H_{0} &= \sum_{j \in B} \Delta \, n_j + \sum_i U \, n_{i\uparrow} n_{i\downarrow} + \frac{1}{2} \sum_{i \neq j} V_{ij} \, n_i n_j 
\end{split} \label{eq:H}
\end{equation}
where the A and B sublattices (colored black and red in the figures) correspond to the anion and cation, respectively. $c^\dagger_i$ creates a hole in the moir\'e valence band with charge $e>0$, and $n_i = n_{i\uparrow} + n_{i\downarrow}$. $\Delta>0$ is  the energy difference of cation and anion orbitals.
For WSe$_2$/WS$_2$,  $\Delta = 14.9$meV is extracted from first-principles band structure calculations \cite{FuChargeTransfer,LiuZhandPrivate}.

\newsec{Trimers}The hopping integrals $t_{ij}$ decrease exponentially as the moir\'e period $L_{\rm M}$ increases.
The on-site and extended two-body interactions $U, V_{ij}>0$ are given by Coulomb integrals in Wannier basis \cite{FuChargeTransfer, LiuZhandPrivate}.
Since $U$ and $V_{ij}$ decrease as power law functions of $L_{\rm M}$, electron-electron interactions dominate over single-particle hopping at large $L_{\rm M}$.
In this strong-coupling regime, the on-site repulsion $U$ is the largest relevant energy scale.
At the filling $n=1$, the system is in an insulating state where all anions are singly occupied and cations unoccupied; i.e., $n^0_i=1$ for $i\in A$ and $n^{0}_i=0$ for $i\in B$.

Upon doping to $n>1$, the additional $n-1$ charges have to occupy the cations in order to avoid the large energy cost of double occupancy. To study the many-body physics at finite doping, we first identify the relevant charged excitations at $n=1$.
For this purpose, it is useful to rewrite $H_0$ in terms of $\delta n_i \equiv n_i -n_i^0 $, the change of occupation relative to the ground state:
\begin{equation}
H_{0} =   \sum_{i \in A} V_a  \delta n_i + \sum_{j \in B} (\Delta + V_c) \delta n_j
      + \frac{1}{2} \sum_{i \neq j} V_{ij} \, \delta n_i \delta n_j. \label{eq:rel H0}
\end{equation}
We have taken the $U=\infty$ limit so that double occupancy is forbidden.
Then, there exist two types of elementary excitations: (1) electrons on the A sublattice of anions with $\delta n = -1$ (charge $-e$), and (2) holes on the B sublattice of cations with $\delta n=1$ (charge $+e$).
$V_c$ ($V_a$) is the self-energy of a hole (electron) due to its electrostatic interaction with all other charges in the ground state, defined by $V_c \; (V_a) =\sum_{j \in A} V_{ij}$ with $i \in B \; (A)$ \cite{Dfoot:Vac}.  The Coulomb energies $V_{ij}$ decrease rapidly due to screening effects when the distance between sites $i$ and $j$ exceeds the distance to nearby metallic gates.

\begin{figure}
  \includegraphics[width=.8\columnwidth]{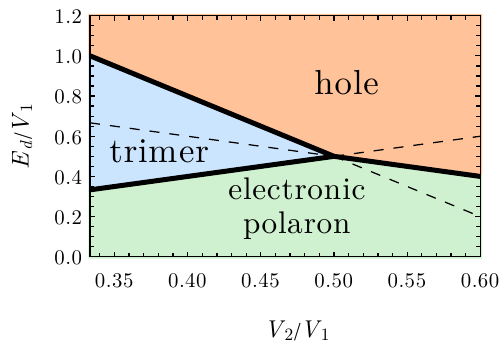}
  \raisebox{1\height}{\includegraphics[width=.18\columnwidth]{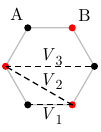}}
  \includegraphics[width=.9\columnwidth]{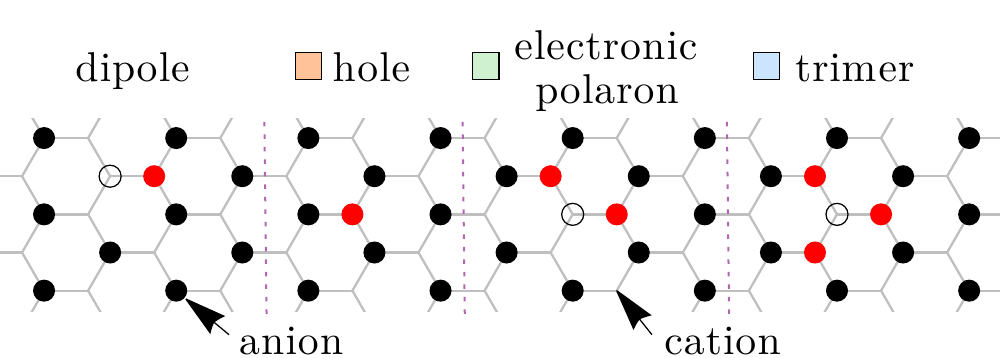}
  \caption{%
    Phase diagram showing which of the three charged excitations (hole, electronic polaron, or trimer) in the charge-transfer insulator at $n=1$ has the lowest energy \emph{per unit charge} 
      as a function of the fundamental gap $E_d$ [defined in \eqnref{eq:Ed}] and the ratio of next-nearest-neighbor to nearest-neighbor repulsion: $V_2/V_1$.
    The dashed lines separate regions where the second lowest energy per charge excitation changes.
    The bottom panel depicts these excitations where filled circles denote a hole,
      while an empty circle denotes a hole that is missing from the charge-transfer insulator ground state.
    Anions and cations are colored black and red, respectively.
    See \figref{fig:detailedExcitations} in the appendix for more information.
  }\label{fig:excitation}
\end{figure}

It follows from \eqnref{eq:rel H0} that adding a hole to the charge-transfer insulator costs energy $E_e - \mu$, where $E_e = \Delta + V_c$ and $\mu$ is the chemical potential for holes. Likewise, adding an electron costs energy $E_{-e} + \mu$ with $E_{-e} = - V_a$. Transferring a charge from an anion to its adjacent cation creates a {\it charge-transfer exciton}, which carries an electric dipole.
Its energy cost $E_d$ is less than the sum of the electron and the hole energies:
\begin{equation}
E_d = E_{e} + E_{-e} - V_1 =\Delta + V_c - V_a - V_1, 
\label{eq:Ed}
\end{equation}
where $V_1$ denotes the nearest-neighbor repulsion (see top-right of \figref{fig:excitation}).
$E_d>0$ defines the fundamental gap of the charge-transfer insulator to local neutral excitations.

Two holes cost energy $ 2 E_e$ when they are spatially separated.
Alternatively, consider binding two adjacent holes with a neutral charge-transfer dipole. The result is a charge-$2e$ composite excitation consisting of three adjacent holes on cations surrounding an electron at the center anion, a ``trimer'' (see bottom of \figref{fig:excitation}).
Its energy cost, written as $E_t = 2E_e - \epsilon_b$, differs from the two separate holes by a pair binding energy $\epsilon_b$:
\begin{eqnarray}
\epsilon_b &=& -E_d + 2V_1 - 3V_2 \nonumber \\
&=& - \Delta + 6V'_2 - 3 V_2 - 3V_3 + \cdots \nonumber \\
&\approx &  - \Delta + 3V_2 - 3V_3 + \cdots, \label{eq:Eb}
\end{eqnarray}
where the second-nearest neighbor interactions $V'_2$ and $V_2$ are within the $A$ and $B$ sublattice respectively, and $\cdots$ denote interactions at larger distance.
In the second and third equalities, we have used \eqnref{eq:Ed} and $V'_2 \approx V_2$.  

Importantly, in a range of realistic material parameters $\Delta$ and $V_n$ (see \figref{fig:excitation}), $\epsilon_b$ is positive so that a charge-$2e$ trimer costs less energy than two individual holes.
The energy gain here comes mostly from the simple fact that pairing two holes into a trimer frees up three second-neighbor pairs in the system,
  which results in the $-3V_2$ energy reduction in \eqnref{eq:Eb}.
It is remarkable that despite the direct mutual repulsion, two doped holes can, at the expense of energy $\Delta$, tightly bind together with a charge-transfer exciton to lower the total electrostatic repulsion energy.

As an example, for slightly twisted WSe$_2$/WS$_2$ with a moir\'e period $L_\text{M}=7$nm and a distance to top and bottom gates equal to $L_\text{M}$,
  a calculation using Wannier functions finds $V_1, V'_2, V_2, V_3 = 1.2998, 0.4599, 0.4780, 0.3239$ in units of $e^2/(\epsilon L_\text{M})$,
  where $\epsilon$ is the permittivity of the dielectric environment \cite{LiuZhandPrivate}.
The Coulomb energies at larger separation are much smaller.
The trimer binding energy is then found to be $\epsilon_b= -14.9 + \frac{72.8}{\epsilon}$ meV.

The finding of charge pairing from Coulomb repulsion in a moir\'e superlattice is our first main result, which forms the basis for our theory at finite doping.
Notably, previous works found an effective attraction between two added charges in small Hubbard-model clusters at intermediate $U/t$ \cite{WhiteMolecules,KivelsonCheckerboard,IsaevOrtiz} or with extended interactions \cite{SlagleKim}.
In this study of extended moir\'e systems, the pair binding energy is already manifest in the strong-coupling limit $t_{ij}=0$
  without explicitly invoking any charge fluctuation or weakly coupled clusters.
Note however, the fact crucial to our analysis that doped charges occupy quantized orbitals localized around discrete lattice points, rather than taking arbitrary positions in the continuum.
It is this quantum-mechanical effect that leads to the quantized energy of trimers.

We also mention in passing that besides holes and trimers, other composite excitations can be energetically favorable in certain parameter ranges.
These include (1) the electronic polaron ($q=e$), which is a bound state of a hole and a dipole, and (2) higher-charge excitations with $q\geq 3e$. By comparing the energy cost of different types of charged excitations at filling $n=1$, we identify the excitation with the least energy per unit charge---the cheapest charge excitation; see \figref{fig:excitation}.

\begin{figure}
  \subfloat[$\delta = 1/7 \approx 0.143$ $q=1,r_1=2,r_2=1$\label{fig:8_7_crystal}]{    \includegraphics[width=.4\columnwidth]{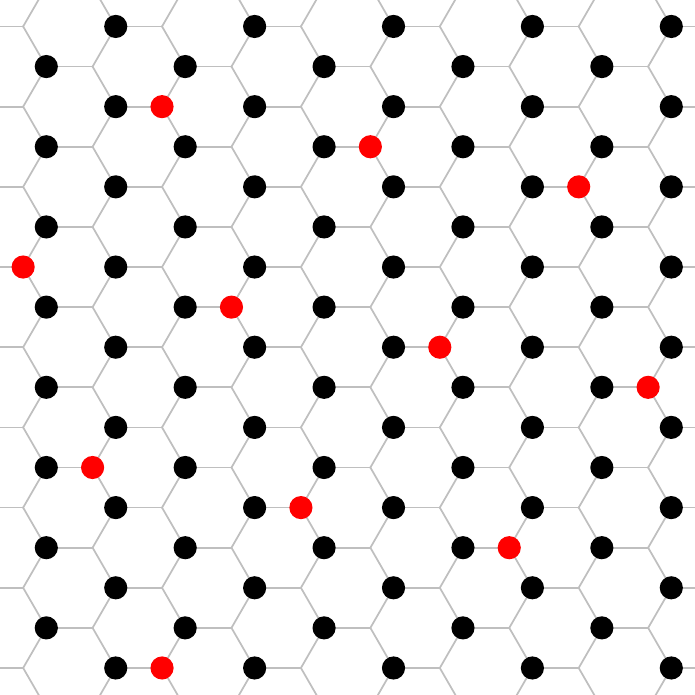}} \hspace{.1\columnwidth}
  \subfloat[$\delta = 1/8 =       0.125$ $q=2,r_1=4,r_2=0$\label{fig:9_8_pairCrystal}]{\includegraphics[width=.4\columnwidth]{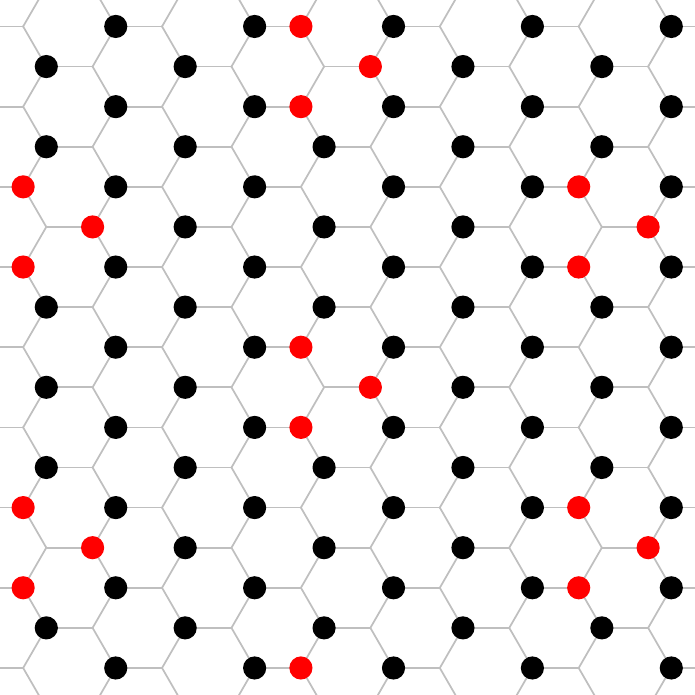}} \\
  \subfloat[$\delta = 1/4 =       0.25 $ $q=1,r_1=2,r_2=0$\label{fig:5_4_crystal}]{    \includegraphics[width=.4\columnwidth]{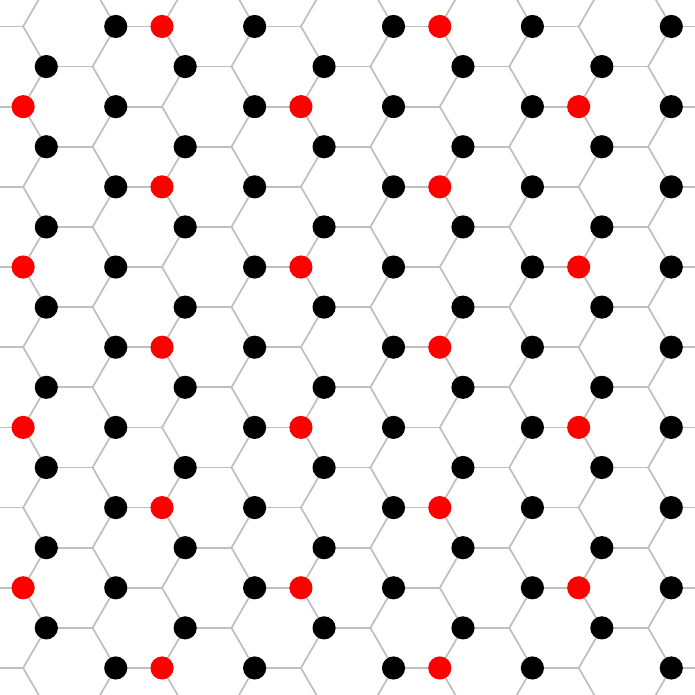}} \hspace{.1\columnwidth}
  \subfloat[$\delta = 2/7 \approx 0.286$ $q=2,r_1=2,r_2=1$\label{fig:9_7_pairCrystal}]{\includegraphics[width=.4\columnwidth]{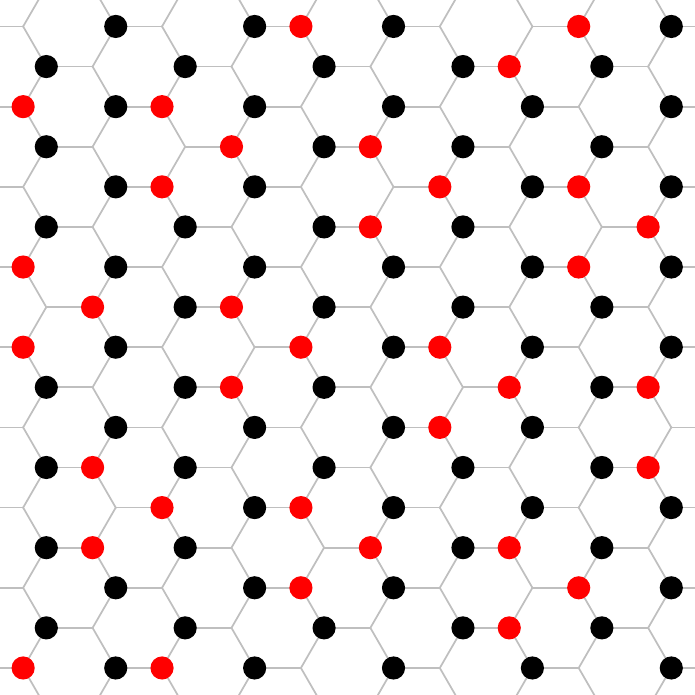}} \\
  \subfloat[$\delta = 1/3 \approx 0.333$ $q=1,r_1=1,r_2=1$\label{fig:4_3_crystal}]{    \includegraphics[width=.4\columnwidth]{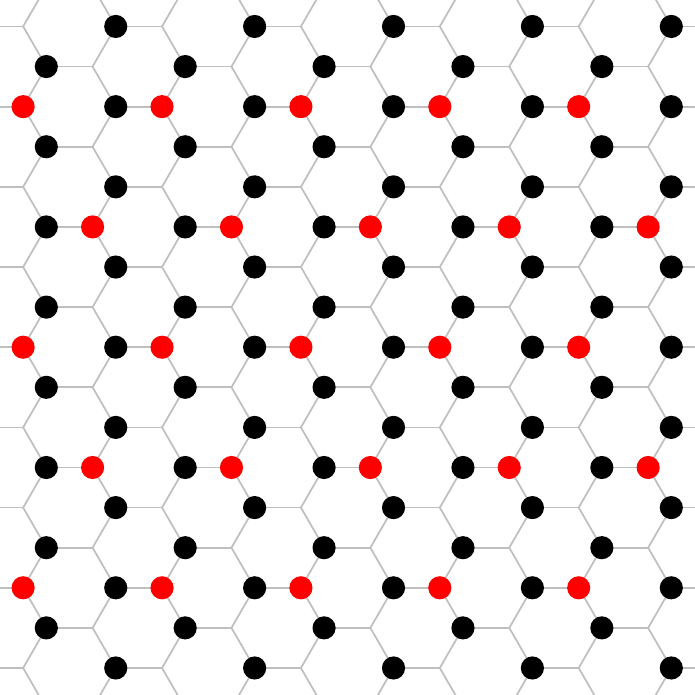}} \hspace{.1\columnwidth}
  \subfloat[$\delta = 1/3 \approx 0.333$                  \label{fig:4_3_pairCrystal}]{\includegraphics[width=.4\columnwidth]{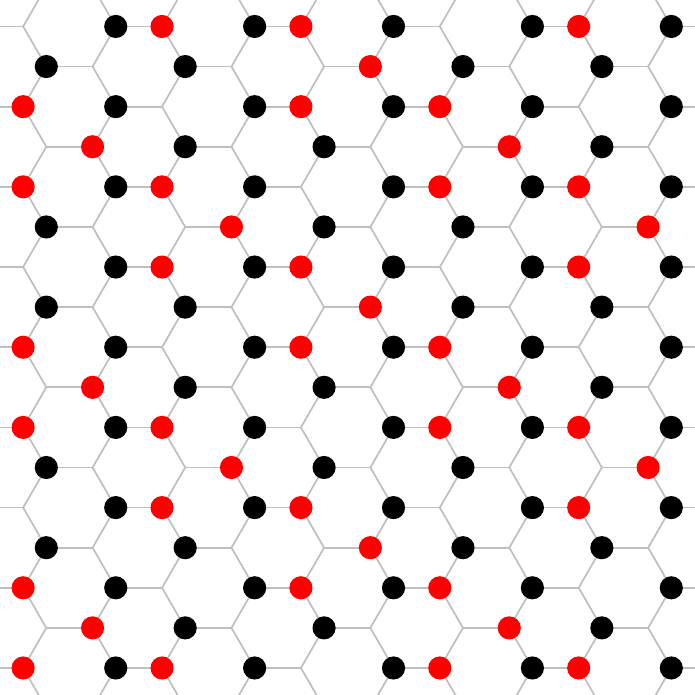}}
  \caption{%
    Examples of commensurate charge (a,c,e) and pair (b,d,f) density waves at various fractional fillings $n=1+\delta$.
    The charge $q$ and integers $r_1$ and $r_2$ from \eqnref{eq:crystal fillings} are listed for the triangular Wigner crystals of holes ($q=e$) and of trimers ($q=2e$).}
  \label{fig:crystals}
\end{figure}

\newsec{Pair Density Waves}Building on these results on few-body excitations, we now study the many-body physics of TMD heterobilayers at fillings $n=1 + \delta$ with small $\delta>0$,
  which are doped charge-transfer insulators.
In particular, we shall develop an analytically-controlled theory to predict pair density waves and superconductivity under appropriate conditions.

For large moir\'e periods where the Coulomb interaction dominates the kinetic energy,
  a periodic array of charge excitations of the least cost is favored.
If the cheapest excitation is a charge-$e$ hole, then the ground state of $H_0$ will be a commensurate charge density wave at e.g. the dopings $\delta = 1/7$, $1/4$, or $1/3$,
  shown in \sfigref{fig:crystals}{a,c,e}.
Similar charge density waves have been discussed in the context of twisted bilayer graphene \cite{PhilipWigner} and extended Hubbard models \cite{HubbardWigner,RademakerChargeOrder,KapciaChargeOrder}.

On the other hand, if the cheapest charge excitation is a charge-$2e$ trimer, then the ground state of $H_0$ is a pair density wave \cite{PairDensityWave} with a commensurate periodicity at dopings such as $\delta=1/8$, $2/7$, or $1/3$, shown in \sfigref{fig:crystals}{b,d,f}.
In particular, the pair density wave at $\delta=2/7$ (shown in \figref{fig:9_7_pairCrystal})
  can be viewed as the closest packing of trimers with negligible inter-trimer interaction (involving only $V_{n\geq5}$),
  in contrast to \figref{fig:4_3_pairCrystal} (which involves $V_{n\geq2}$).
Possible evidence of such a state can be seen from the Berkely group's recent data;
  see \figref{fig:BerkeleyData} of the appendix.
Furthermore, possible evidence for one of the $\delta=1/3$ states was observed in \refcite{ShanTMD}.

More generally, we predict that at low temperature, clean TMD heterobilayers with a large moir\'e period should exhibit a sequence of insulating density wave states at the following fillings when a commensurate triangular lattice of charge excitations is formed:
\begin{align}
 n &= 1 + \frac{q}{r_1^2+r_1 r_2+r_2^2} & \mathord{\vcenter{\hbox{\includegraphics[scale=.5]{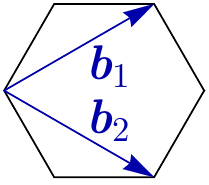}}}} &
  \label{eq:crystal fillings}.
\end{align}
The integer $q>0$ is the charge per excitation, and the integers $r_i\geq0$ specify the Bravais lattice vector $r_1 \mathbf b_1 + r_2 \mathbf b_2$ of the commensurate density wave, with
$\mathbf b_1, \mathbf b_2$ denoting the two lattice vectors of the moir\'e honeycomb lattice shown in \eqnref{eq:crystal fillings}.


\newsec{Resonantly-Paired Superconductivity}While Coulomb repulsion favors density waves, the single-particle hopping term $H_K$ favors charge delocalization. In the following, we study the competition between Coulomb energy and kinetic energy in the interesting and experimentally relevant parameter regime $\epsilon_b>0$, where the trimer has a lower energy than two separated holes in the limit $t_{ij} = 0$ (the trimer region of \figref{fig:excitation}). For simplicity, we consider the scenario in which the system is fully spin polarized, which is experimentally realized in WSe$_2$/WS$_2$ under a small magnetic field (less than $1$T at $1.6$K) \cite{FaiTMD}.
Note that an applied magnetic field will not weaken charge pairing since charge pairing occurs through electrostatic interactions and without double occupancy.

\begin{figure}[t!]
  \subfloat[pair density wave \mbox{$\quad\;\;$or BEC}\label{fig:BEC}]{\includegraphics[width=.49\columnwidth]{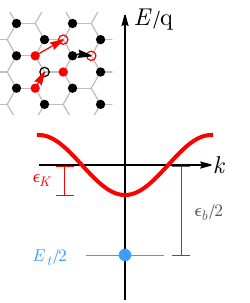}} 
  \subfloat[resonantly-paired \mbox{$\quad\;\;\,$superconductivity}\label{fig:resonance}]{\includegraphics[width=.49\columnwidth]{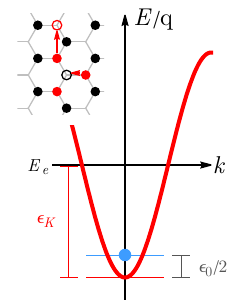}} \\
  \caption{%
  Energy spectrum of the dispersive hole band and the trimer.
  (a) When the trimer energy per charge, $E_t/2$, lies below the hole band bottom,
    a pair density wave (PDW) or Bose-Einstein condensate of trimers is expected.
  (inset) A third-order trimer hopping process. 
  (b) When $E_t/2$ lies close to the hole band bottom,
   low-energy holes interact resonantly with trimers, leading to superconductivity.
  (inset) A second-order process in which a trimer decays into two holes separated by a distance $2L_\text{M}$.
  }\label{fig:mixture}
\end{figure}

Single-particle hopping $t_{ij}$ between cation sites on the triangular sublattice leads to a dispersive band of doped holes.
Then, the lowest-energy hole excitation is in the delocalized state at the bottom of this band.
Its energy is $E_e -\epsilon_K$, where $\epsilon_K>0$ is proportional to the hopping integral $t$. In contrast, as a composite excitation,  the trimer can only hop via a high-order process involving high-energy intermediate states (see \figref{fig:BEC}).
In the strong-coupling regime, 
the trimer hopping integral is on the order of $t^3/V^2$ with $V\sim e^2/\epsilon L_\text{M}$ and thus likely negligible.

With $t_{ij} \neq 0$, it is important to consider the hybridization between trimers and holes. This occurs when a constituent $e$ charge in a trimer hops back to the center anion, leaving behind two adjacent holes in a high-energy state due to their strong mutual repulsion $V_2$. To lower their energy, these two remaining holes tend to hop away from each other.
Thus, converting a trimer into two weakly-interacting holes that are sufficiently apart requires at least a second-order process, shown in \figref{fig:resonance}.
Due to the large energy barrier in the pathway between the trimer and two free holes, the trimer remains a long-lived quasi-bound state.

Therefore, at finite doping, we are faced with a mixture of charge-$e$ holes and charge-$2e$ trimers that are hybridized and at a total charge density $\delta$.
For small doping $\delta$, the typical distance between doped charges is much larger than the moir\'e period $L_\text{M}$;
  hence the underlying moir\'e lattice only plays a minor role.
Thus, the essential low-energy physics is captured by a boson-fermion model in the continuum, which we introduce for doped charge-transfer insulators:
\begin{widetext}
\begin{align}
\begin{split}
 &  H_{\delta} = \int\! d {\bf r} \; \sum_{\sigma=\pm} \psi^\dagger_\sigma \!\left(-\frac{\nabla^2}{2m}\right)\! \psi_\sigma + \epsilon_0 \, \phi^\dagger \phi
   + g\, (\phi \psi^\dagger_{+} \psi^\dagger_-  + \phi^\dagger \psi_- \psi_+)
  - \mu \,  n({\bf r})
  + \frac{1}{2} \int\! d {\bf r'} \; V(|{\bf r - \bf r'}|) \,  n({\bf r})  n({\bf r'}) \\
 & \textrm{with } n({\bf r}) = \sum_{\sigma=\pm} \psi^\dagger_\sigma \psi_\sigma + 2 \phi^\dagger \phi 
\end{split} \label{eq:Hdelta}
\end{align}
\end{widetext}
where $\psi$ and $\phi$ denote the itinerant hole and immobile trimer, respectively.

According to band structure calculations for WSe$_2$/WS$_2$, $t_{ij}>0$ between nearest-neighbor cations.
Hence the band of doped holes has two degenerate minima at corners of the moir\'e Brillouin zone
$\pm {\bf K} =
(0,\pm\frac{4\pi}{3L_\text{M}})$ \cite{FuChargeTransfer},
which are denoted by the valley index $\sigma = \pm$ in \eqnref{eq:Hdelta}. $m \propto 1/t$ is the effective mass at the band bottom.
$\epsilon_0 = -\epsilon_b + 2\epsilon_K$ with $\epsilon_K \propto t$ is the energy difference between a trimer and two delocalized holes at the band bottom.
Since the hopping integral $t$ changes significantly with the moir\'e period,  the detuning parameter $\epsilon_0$ is tunable by varying the twist angle.

Two types of interactions are included in our model Hamiltonian $H_\delta$ and play  dominant roles in the {\it low density} regime $\delta \ll 1$:
  (1) the extended Coulomb interaction $V(r)$, whose range is determined by the distance to metallic gates, and
  (2) the local hybridization $g$ between holes and trimers.
The form of the hybridization is dictated by symmetry.
The trimer state with maximal spin is invariant under three-fold rotation around the center anion and odd under reflection which exchanges a pair of holes (fermions).
Therefore, the trimer hybridizes with a valley-singlet pair of holes  $\psi^\dagger_+ \psi^\dagger_-$, which transforms in the same way
  (note that reflection interchanges $\pm \bf K$).
Despite being a weak local interaction, the hole-trimer hybridization can have dramatic consequences for our system at low density.

Our model exhibits an enormously rich phase diagram resulting from the interplay of (1) the kinetic energy of holes,  (2) the binding energy of trimers, (3) Coulomb repulsion between charges, and (4) the hybridization between holes and trimers.
In particular, we show that pair density wave and superconducting ground states emerge in certain parameter regimes.

If the hopping integral $t$ in the microscopic Hamiltonian \eqref{eq:H} is small compared to the binding energy of trimers $\epsilon_b>0$ (see \figref{fig:BEC}), then the kinetic energy of holes is of minor importance.
Therefore $\epsilon_0\approx -\epsilon_b<0$, and doped charges go into trimers.
At finite charge density up to $\delta =1/3$, the (screened) Coulomb repulsion $V(r)$ between charges leads to a Wigner crystal of charge-$2e$ trimers, which takes a triangular lattice structure.
At the dopings specified by \eqnref{eq:crystal fillings}, this trimer Wigner crystal is commensurate with and pinned by the moir\'e potential. The resulting state is a gapped and insulating pair density wave. 
At sufficiently low doping where the average distance between trimers exceeds the range of $V(r)$,
  the density wave state becomes fragile and potentially unstable to Bose condensation of trimers when their small hopping amplitudes are taken into account.

As the hopping integral $t$ increases, the bottom of the hole band is lowered and eventually falls below $E_t/2$, as shown in \figref{fig:resonance}.
Correspondingly, the bare detuning parameter $\epsilon_0$ changes from negative to positive.
The true detuning parameter $\epsilon$ is renormalized by the hole-trimer hybridization: $\epsilon=\epsilon_0 - o(g^2)$ \cite{Cfoot:renormalize}.
At negative detuning $\epsilon<0$, there exists a true bound state of two $e$ charges, which is a superposition of a trimer and a cloud of two holes.
At positive detuning $\epsilon>0$, no such bound state exists.
However, when the detuning is small, low-energy holes and trimers are strongly hybridized.
Two $e$ charges at low energy or large de Broglie wavelength scatter {\it resonantly} in the $s$-wave valley-singlet channel.
Such resonant interaction is universally characterized by a large scattering length, which can exceed the range of screened Coulomb repulsion $V(r)$. The scattering length
is positive (negative)  for $\epsilon>0$ ($\epsilon<0$) and diverges at $\epsilon=0$.



The physics of the resonant interaction via the trimer channel is reminiscent of a Feshbach resonance in an ultracold Fermi gas \cite{GurarieRadzihovskyFeshbach}.
Under the resonant condition, the ground state of our system at low doping is a superconductor with same-spin pairing in the valley-singlet channel (corresponding to $f$-wave pairing symmetry on the honeycomb lattice).
A crossover between the Bose-Einstein condensate (BEC) and Bardeen-Cooper-Schrieffer (BCS) regimes \cite{BECBCS} can be achieved as a function of doping density $\delta$ and the detuning $\epsilon$. See \appref{app:details} for a mean-field theory analysis and more microscopic details.

It should be noted that the origin of the resonant interaction in our system is completely different from the case of cold atoms.
Unlike the molecule formed by two atoms in empty space, the trimer is a {\it charge-transfer excitation} in a {\it many-body} ``vacuum'' at the filling $n=1$.
It is remarkable that spin-polarized superconductivity can be realized in a solid-state system with purely Coulomb repulsion.
We emphasize that our prediction of superconductivity is rigorous in the regime of
  weak kinetic energy ($t \ll V$), polarized spins, small doping ($\delta \ll 1$), and resonance (small $\epsilon_0$).
These limits can be reasonably achieved in moir\'e TMD systems.
It will be interesting to consider relaxing some of these assumptions (e.g. polarized spins) in future work
  since other possibilities are likely outside of this regime.


Note that these limits are not as applicable to cuprates, which are antiferromagnetic at zero doping and become superconducting only under a sufficient doping after antiferromagnetism is destroyed.

\newsec{Discussion}To summarize, we developed a strong-coupling theory to predict electron pairing from repulsion via charge-transfer excitations in TMD heterobilayers.
We further predict insulating pair density wave states at a sequence of doping levels.
Finally, we show that with the increase of electron itinerancy, the resonant interaction between itinerant holes and local charge-$2e$ pairs leads to unconventional superconductivity.
Since our theory is based on a general doped charge-transfer insulator,
  our theory is broadly applicable to other charge-transfer insulating moir\'e materials,
  such as MoS$_2$/MoS$_2$ and other TMD homobilayers in an electric field \cite{FuHomobilayer,MacDonaldHomobilayer,TMDRealization}.
Charge-transfer physics occurs when $U > \Delta$,
  where $U$ is large due to localized orbitals in moir\'e TMD \cite{CrommieImaging}
  and $\Delta$ can be arbitrarily small in TMD homobilayers where it is proportional to the applied electric field.
Our pairing mechanism may also shed new insight on other moir\'e materials, such as twisted graphene multilayers,
  where charge redistribution under doping may be important \cite{RademakerTBG,GuineaTGB} and spin-polarized superconductivity may have been observed \cite{KimPolarizedSC,WangPolarizedSC}.
We hope our prediction of new fascinating correlated states in moir\'e materials can stimulate further activity and find experimental realization soon.

\begin{acknowledgments}
We are grateful to Steve Kivelson for a stimulating and insightful discussion.
We thank Jason Alicea, Chenhao Jin, Patrick Lee, Stevan Nadj-Perge, Gil Refael, Alex Thomson, and Sagar Vijay for helpful conversations.
We thank Yang Zhang, Noah Yuan, and especially Tongtong Liu for collaboration on related works, and Kin Fai Mak, Jie Shan, and Feng Wang for valuable discussions on experiments.
It is our pleasure to thank the organizers of the KITP conference ``Topological Quantum Matter: From Fantasy to Reality,'' where this work was initiated.
K.S. is supported by the Walter Burke Institute for Theoretical Physics at Caltech.
L.F. is supported by the DOE Office of Basic Energy Sciences under Award DE-SC0018945, and in part by a Simons Investigator award from the Simons Foundation.
\end{acknowledgments}

\bibliography{TMD}

\begin{thebibliography}{65}%
\makeatletter
\providecommand \@ifxundefined [1]{%
 \@ifx{#1\undefined}
}%
\providecommand \@ifnum [1]{%
 \ifnum #1\expandafter \@firstoftwo
 \else \expandafter \@secondoftwo
 \fi
}%
\providecommand \@ifx [1]{%
 \ifx #1\expandafter \@firstoftwo
 \else \expandafter \@secondoftwo
 \fi
}%
\providecommand \natexlab [1]{#1}%
\providecommand \enquote  [1]{``#1''}%
\providecommand \bibnamefont  [1]{#1}%
\providecommand \bibfnamefont [1]{#1}%
\providecommand \citenamefont [1]{#1}%
\providecommand \href@noop [0]{\@secondoftwo}%
\providecommand \href [0]{\begingroup \@sanitize@url \@href}%
\providecommand \@href[1]{\@@startlink{#1}\@@href}%
\providecommand \@@href[1]{\endgroup#1\@@endlink}%
\providecommand \@sanitize@url [0]{\catcode `\\12\catcode `\$12\catcode
  `\&12\catcode `\#12\catcode `\^12\catcode `\_12\catcode `\%12\relax}%
\providecommand \@@startlink[1]{}%
\providecommand \@@endlink[0]{}%
\providecommand \url  [0]{\begingroup\@sanitize@url \@url }%
\providecommand \@url [1]{\endgroup\@href {#1}{\urlprefix }}%
\providecommand \urlprefix  [0]{URL }%
\providecommand \Eprint [0]{\href }%
\providecommand \doibase [0]{http://dx.doi.org/}%
\providecommand \selectlanguage [0]{\@gobble}%
\providecommand \bibinfo  [0]{\@secondoftwo}%
\providecommand \bibfield  [0]{\@secondoftwo}%
\providecommand \translation [1]{[#1]}%
\providecommand \BibitemOpen [0]{}%
\providecommand \bibitemStop [0]{}%
\providecommand \bibitemNoStop [0]{.\EOS\space}%
\providecommand \EOS [0]{\spacefactor3000\relax}%
\providecommand \BibitemShut  [1]{\csname bibitem#1\endcsname}%
\let\auto@bib@innerbib\@empty
\bibitem [{\citenamefont {{Bistritzer}}\ and\ \citenamefont
  {{MacDonald}}(2011)}]{MacDonaldGraphene}%
  \BibitemOpen
  \bibfield  {author} {\bibinfo {author} {\bibfnamefont {Rafi}\ \bibnamefont
  {{Bistritzer}}}\ and\ \bibinfo {author} {\bibfnamefont {Allan~H.}\
  \bibnamefont {{MacDonald}}},\ }\bibfield  {title} {\enquote {\bibinfo {title}
  {{Moir{\'e} bands in twisted double-layer graphene}},}\ }\href {\doibase
  10.1073/pnas.1108174108} {\bibfield  {journal} {\bibinfo  {journal}
  {Proceedings of the National Academy of Science}\ }\textbf {\bibinfo {volume}
  {108}},\ \bibinfo {pages} {12233--12237} (\bibinfo {year} {2011})},\ \Eprint
  {http://arxiv.org/abs/1009.4203} {arXiv:1009.4203} \BibitemShut {NoStop}%
\bibitem [{\citenamefont {{Balents}}(2019)}]{BalentsMoire}%
  \BibitemOpen
  \bibfield  {author} {\bibinfo {author} {\bibfnamefont {Leon}\ \bibnamefont
  {{Balents}}},\ }\bibfield  {title} {\enquote {\bibinfo {title} {{General
  continuum model for twisted bilayer graphene and arbitrary smooth
  deformations}},}\ }\href {\doibase 10.21468/SciPostPhys.7.4.048} {\bibfield
  {journal} {\bibinfo  {journal} {SciPost Physics}\ }\textbf {\bibinfo {volume}
  {7}},\ \bibinfo {eid} {048} (\bibinfo {year} {2019})},\ \Eprint
  {http://arxiv.org/abs/1909.01545} {arXiv:1909.01545} \BibitemShut {NoStop}%
\bibitem [{\citenamefont {{Lopes Dos Santos}}\ \emph
  {et~al.}(2007)\citenamefont {{Lopes Dos Santos}}, \citenamefont {{Peres}},\
  and\ \citenamefont {{Castro Neto}}}]{CastroTwist}%
  \BibitemOpen
  \bibfield  {author} {\bibinfo {author} {\bibfnamefont {J.~M.~B.}\
  \bibnamefont {{Lopes Dos Santos}}}, \bibinfo {author} {\bibfnamefont
  {N.~M.~R.}\ \bibnamefont {{Peres}}}, \ and\ \bibinfo {author} {\bibfnamefont
  {A.~H.}\ \bibnamefont {{Castro Neto}}},\ }\bibfield  {title} {\enquote
  {\bibinfo {title} {{Graphene Bilayer with a Twist: Electronic Structure}},}\
  }\href {\doibase 10.1103/PhysRevLett.99.256802} {\bibfield  {journal}
  {\bibinfo  {journal} {\prl}\ }\textbf {\bibinfo {volume} {99}},\ \bibinfo
  {eid} {256802} (\bibinfo {year} {2007})},\ \Eprint
  {http://arxiv.org/abs/0704.2128} {arXiv:0704.2128} \BibitemShut {NoStop}%
\bibitem [{\citenamefont {{Cao}}\ \emph
  {et~al.}(2018{\natexlab{a}})\citenamefont {{Cao}}, \citenamefont {{Fatemi}},
  \citenamefont {{Fang}}, \citenamefont {{Watanabe}}, \citenamefont
  {{Taniguchi}}, \citenamefont {{Kaxiras}},\ and\ \citenamefont
  {{Jarillo-Herrero}}}]{Pablo2018}%
  \BibitemOpen
  \bibfield  {author} {\bibinfo {author} {\bibfnamefont {Yuan}\ \bibnamefont
  {{Cao}}}, \bibinfo {author} {\bibfnamefont {Valla}\ \bibnamefont {{Fatemi}}},
  \bibinfo {author} {\bibfnamefont {Shiang}\ \bibnamefont {{Fang}}}, \bibinfo
  {author} {\bibfnamefont {Kenji}\ \bibnamefont {{Watanabe}}}, \bibinfo
  {author} {\bibfnamefont {Takashi}\ \bibnamefont {{Taniguchi}}}, \bibinfo
  {author} {\bibfnamefont {Efthimios}\ \bibnamefont {{Kaxiras}}}, \ and\
  \bibinfo {author} {\bibfnamefont {Pablo}\ \bibnamefont {{Jarillo-Herrero}}},\
  }\bibfield  {title} {\enquote {\bibinfo {title} {{Unconventional
  superconductivity in magic-angle graphene superlattices}},}\ }\href {\doibase
  10.1038/nature26160} {\bibfield  {journal} {\bibinfo  {journal} {\nat}\
  }\textbf {\bibinfo {volume} {556}},\ \bibinfo {pages} {43--50} (\bibinfo
  {year} {2018}{\natexlab{a}})},\ \Eprint {http://arxiv.org/abs/1803.02342}
  {arXiv:1803.02342} \BibitemShut {NoStop}%
\bibitem [{\citenamefont {{Cao}}\ \emph
  {et~al.}(2018{\natexlab{b}})\citenamefont {{Cao}}, \citenamefont {{Fatemi}},
  \citenamefont {{Demir}}, \citenamefont {{Fang}}, \citenamefont {{Tomarken}},
  \citenamefont {{Luo}}, \citenamefont {{Sanchez-Yamagishi}}, \citenamefont
  {{Watanabe}}, \citenamefont {{Taniguchi}}, \citenamefont {{Kaxiras}},
  \citenamefont {{Ashoori}},\ and\ \citenamefont
  {{Jarillo-Herrero}}}]{PabloInsulator}%
  \BibitemOpen
  \bibfield  {author} {\bibinfo {author} {\bibfnamefont {Yuan}\ \bibnamefont
  {{Cao}}}, \bibinfo {author} {\bibfnamefont {Valla}\ \bibnamefont {{Fatemi}}},
  \bibinfo {author} {\bibfnamefont {Ahmet}\ \bibnamefont {{Demir}}}, \bibinfo
  {author} {\bibfnamefont {Shiang}\ \bibnamefont {{Fang}}}, \bibinfo {author}
  {\bibfnamefont {Spencer~L.}\ \bibnamefont {{Tomarken}}}, \bibinfo {author}
  {\bibfnamefont {Jason~Y.}\ \bibnamefont {{Luo}}}, \bibinfo {author}
  {\bibfnamefont {Javier~D.}\ \bibnamefont {{Sanchez-Yamagishi}}}, \bibinfo
  {author} {\bibfnamefont {Kenji}\ \bibnamefont {{Watanabe}}}, \bibinfo
  {author} {\bibfnamefont {Takashi}\ \bibnamefont {{Taniguchi}}}, \bibinfo
  {author} {\bibfnamefont {Efthimios}\ \bibnamefont {{Kaxiras}}}, \bibinfo
  {author} {\bibfnamefont {Ray~C.}\ \bibnamefont {{Ashoori}}}, \ and\ \bibinfo
  {author} {\bibfnamefont {Pablo}\ \bibnamefont {{Jarillo-Herrero}}},\
  }\bibfield  {title} {\enquote {\bibinfo {title} {{Correlated insulator
  behaviour at half-filling in magic-angle graphene superlattices}},}\ }\href
  {\doibase 10.1038/nature26154} {\bibfield  {journal} {\bibinfo  {journal}
  {\nat}\ }\textbf {\bibinfo {volume} {556}},\ \bibinfo {pages} {80--84}
  (\bibinfo {year} {2018}{\natexlab{b}})},\ \Eprint
  {http://arxiv.org/abs/1802.00553} {arXiv:1802.00553} \BibitemShut {NoStop}%
\bibitem [{\citenamefont {{Lu}}\ \emph {et~al.}(2019)\citenamefont {{Lu}},
  \citenamefont {{Stepanov}}, \citenamefont {{Yang}}, \citenamefont {{Xie}},
  \citenamefont {{Aamir}}, \citenamefont {{Das}}, \citenamefont {{Urgell}},
  \citenamefont {{Watanabe}}, \citenamefont {{Taniguchi}}, \citenamefont
  {{Zhang}}, \citenamefont {{Bachtold}}, \citenamefont {{MacDonald}},\ and\
  \citenamefont {{Efetov}}}]{Dmitri2019}%
  \BibitemOpen
  \bibfield  {author} {\bibinfo {author} {\bibfnamefont {Xiaobo}\ \bibnamefont
  {{Lu}}}, \bibinfo {author} {\bibfnamefont {Petr}\ \bibnamefont {{Stepanov}}},
  \bibinfo {author} {\bibfnamefont {Wei}\ \bibnamefont {{Yang}}}, \bibinfo
  {author} {\bibfnamefont {Ming}\ \bibnamefont {{Xie}}}, \bibinfo {author}
  {\bibfnamefont {Mohammed~Ali}\ \bibnamefont {{Aamir}}}, \bibinfo {author}
  {\bibfnamefont {Ipsita}\ \bibnamefont {{Das}}}, \bibinfo {author}
  {\bibfnamefont {Carles}\ \bibnamefont {{Urgell}}}, \bibinfo {author}
  {\bibfnamefont {Kenji}\ \bibnamefont {{Watanabe}}}, \bibinfo {author}
  {\bibfnamefont {Takashi}\ \bibnamefont {{Taniguchi}}}, \bibinfo {author}
  {\bibfnamefont {Guangyu}\ \bibnamefont {{Zhang}}}, \bibinfo {author}
  {\bibfnamefont {Adrian}\ \bibnamefont {{Bachtold}}}, \bibinfo {author}
  {\bibfnamefont {Allan~H.}\ \bibnamefont {{MacDonald}}}, \ and\ \bibinfo
  {author} {\bibfnamefont {Dmitri~K.}\ \bibnamefont {{Efetov}}},\ }\bibfield
  {title} {\enquote {\bibinfo {title} {{Superconductors, orbital magnets and
  correlated states in magic-angle bilayer graphene}},}\ }\href {\doibase
  10.1038/s41586-019-1695-0} {\bibfield  {journal} {\bibinfo  {journal} {\nat}\
  }\textbf {\bibinfo {volume} {574}},\ \bibinfo {pages} {653--657} (\bibinfo
  {year} {2019})},\ \Eprint {http://arxiv.org/abs/1903.06513}
  {arXiv:1903.06513} \BibitemShut {NoStop}%
\bibitem [{\citenamefont {{Chen}}\ \emph
  {et~al.}(2019{\natexlab{a}})\citenamefont {{Chen}}, \citenamefont {{Jiang}},
  \citenamefont {{Wu}}, \citenamefont {{Lyu}}, \citenamefont {{Li}},
  \citenamefont {{Chittari}}, \citenamefont {{Watanabe}}, \citenamefont
  {{Taniguchi}}, \citenamefont {{Shi}}, \citenamefont {{Jung}}, \citenamefont
  {{Zhang}},\ and\ \citenamefont {{Wang}}}]{Wang2018}%
  \BibitemOpen
  \bibfield  {author} {\bibinfo {author} {\bibfnamefont {Guorui}\ \bibnamefont
  {{Chen}}}, \bibinfo {author} {\bibfnamefont {Lili}\ \bibnamefont {{Jiang}}},
  \bibinfo {author} {\bibfnamefont {Shuang}\ \bibnamefont {{Wu}}}, \bibinfo
  {author} {\bibfnamefont {Bosai}\ \bibnamefont {{Lyu}}}, \bibinfo {author}
  {\bibfnamefont {Hongyuan}\ \bibnamefont {{Li}}}, \bibinfo {author}
  {\bibfnamefont {Bheema~Lingam}\ \bibnamefont {{Chittari}}}, \bibinfo {author}
  {\bibfnamefont {Kenji}\ \bibnamefont {{Watanabe}}}, \bibinfo {author}
  {\bibfnamefont {Takashi}\ \bibnamefont {{Taniguchi}}}, \bibinfo {author}
  {\bibfnamefont {Zhiwen}\ \bibnamefont {{Shi}}}, \bibinfo {author}
  {\bibfnamefont {Jeil}\ \bibnamefont {{Jung}}}, \bibinfo {author}
  {\bibfnamefont {Yuanbo}\ \bibnamefont {{Zhang}}}, \ and\ \bibinfo {author}
  {\bibfnamefont {Feng}\ \bibnamefont {{Wang}}},\ }\bibfield  {title} {\enquote
  {\bibinfo {title} {{Evidence of a gate-tunable Mott insulator in a trilayer
  graphene moir{\'e} superlattice}},}\ }\href {\doibase
  10.1038/s41567-018-0387-2} {\bibfield  {journal} {\bibinfo  {journal} {Nature
  Physics}\ }\textbf {\bibinfo {volume} {15}},\ \bibinfo {pages} {237--241}
  (\bibinfo {year} {2019}{\natexlab{a}})},\ \Eprint
  {http://arxiv.org/abs/1803.01985} {arXiv:1803.01985} \BibitemShut {NoStop}%
\bibitem [{\citenamefont {{Choi}}\ \emph {et~al.}(2019)\citenamefont {{Choi}},
  \citenamefont {{Kemmer}}, \citenamefont {{Peng}}, \citenamefont {{Thomson}},
  \citenamefont {{Arora}}, \citenamefont {{Polski}}, \citenamefont {{Zhang}},
  \citenamefont {{Ren}}, \citenamefont {{Alicea}}, \citenamefont {{Refael}},
  \citenamefont {{von Oppen}}, \citenamefont {{Watanabe}}, \citenamefont
  {{Taniguchi}},\ and\ \citenamefont {{Nadj-Perge}}}]{NadjPerge2019}%
  \BibitemOpen
  \bibfield  {author} {\bibinfo {author} {\bibfnamefont {Youngjoon}\
  \bibnamefont {{Choi}}}, \bibinfo {author} {\bibfnamefont {Jeannette}\
  \bibnamefont {{Kemmer}}}, \bibinfo {author} {\bibfnamefont {Yang}\
  \bibnamefont {{Peng}}}, \bibinfo {author} {\bibfnamefont {Alex}\ \bibnamefont
  {{Thomson}}}, \bibinfo {author} {\bibfnamefont {Harpreet}\ \bibnamefont
  {{Arora}}}, \bibinfo {author} {\bibfnamefont {Robert}\ \bibnamefont
  {{Polski}}}, \bibinfo {author} {\bibfnamefont {Yiran}\ \bibnamefont
  {{Zhang}}}, \bibinfo {author} {\bibfnamefont {Hechen}\ \bibnamefont {{Ren}}},
  \bibinfo {author} {\bibfnamefont {Jason}\ \bibnamefont {{Alicea}}}, \bibinfo
  {author} {\bibfnamefont {Gil}\ \bibnamefont {{Refael}}}, \bibinfo {author}
  {\bibfnamefont {Felix}\ \bibnamefont {{von Oppen}}}, \bibinfo {author}
  {\bibfnamefont {Kenji}\ \bibnamefont {{Watanabe}}}, \bibinfo {author}
  {\bibfnamefont {Takashi}\ \bibnamefont {{Taniguchi}}}, \ and\ \bibinfo
  {author} {\bibfnamefont {Stevan}\ \bibnamefont {{Nadj-Perge}}},\ }\bibfield
  {title} {\enquote {\bibinfo {title} {{Electronic correlations in twisted
  bilayer graphene near the magic angle}},}\ }\href {\doibase
  10.1038/s41567-019-0606-5} {\bibfield  {journal} {\bibinfo  {journal} {Nature
  Physics}\ }\textbf {\bibinfo {volume} {15}},\ \bibinfo {pages} {1174--1180}
  (\bibinfo {year} {2019})},\ \Eprint {http://arxiv.org/abs/1901.02997}
  {arXiv:1901.02997} \BibitemShut {NoStop}%
\bibitem [{\citenamefont {{Shen}}\ \emph {et~al.}(2020)\citenamefont {{Shen}},
  \citenamefont {{Chu}}, \citenamefont {{Wu}}, \citenamefont {{Li}},
  \citenamefont {{Wang}}, \citenamefont {{Zhao}}, \citenamefont {{Tang}},
  \citenamefont {{Liu}}, \citenamefont {{Tian}}, \citenamefont {{Watanabe}},
  \citenamefont {{Taniguchi}}, \citenamefont {{Yang}}, \citenamefont {{Meng}},
  \citenamefont {{Shi}}, \citenamefont {{Yazyev}},\ and\ \citenamefont
  {{Zhang}}}]{Zhang2019}%
  \BibitemOpen
  \bibfield  {author} {\bibinfo {author} {\bibfnamefont {Cheng}\ \bibnamefont
  {{Shen}}}, \bibinfo {author} {\bibfnamefont {Yanbang}\ \bibnamefont {{Chu}}},
  \bibinfo {author} {\bibfnamefont {QuanSheng}\ \bibnamefont {{Wu}}}, \bibinfo
  {author} {\bibfnamefont {Na}~\bibnamefont {{Li}}}, \bibinfo {author}
  {\bibfnamefont {Shuopei}\ \bibnamefont {{Wang}}}, \bibinfo {author}
  {\bibfnamefont {Yanchong}\ \bibnamefont {{Zhao}}}, \bibinfo {author}
  {\bibfnamefont {Jian}\ \bibnamefont {{Tang}}}, \bibinfo {author}
  {\bibfnamefont {Jieying}\ \bibnamefont {{Liu}}}, \bibinfo {author}
  {\bibfnamefont {Jinpeng}\ \bibnamefont {{Tian}}}, \bibinfo {author}
  {\bibfnamefont {Kenji}\ \bibnamefont {{Watanabe}}}, \bibinfo {author}
  {\bibfnamefont {Takashi}\ \bibnamefont {{Taniguchi}}}, \bibinfo {author}
  {\bibfnamefont {Rong}\ \bibnamefont {{Yang}}}, \bibinfo {author}
  {\bibfnamefont {Zi~Yang}\ \bibnamefont {{Meng}}}, \bibinfo {author}
  {\bibfnamefont {Dongxia}\ \bibnamefont {{Shi}}}, \bibinfo {author}
  {\bibfnamefont {Oleg~V.}\ \bibnamefont {{Yazyev}}}, \ and\ \bibinfo {author}
  {\bibfnamefont {Guangyu}\ \bibnamefont {{Zhang}}},\ }\bibfield  {title}
  {\enquote {\bibinfo {title} {{Correlated states in twisted double bilayer
  graphene}},}\ }\href {\doibase 10.1038/s41567-020-0825-9} {\bibfield
  {journal} {\bibinfo  {journal} {Nature Physics}\ }\textbf {\bibinfo {volume}
  {16}},\ \bibinfo {pages} {520--525} (\bibinfo {year} {2020})},\ \Eprint
  {http://arxiv.org/abs/1903.06952} {arXiv:1903.06952} \BibitemShut {NoStop}%
\bibitem [{\citenamefont {{Carr}}\ \emph {et~al.}(2019)\citenamefont {{Carr}},
  \citenamefont {{Fang}}, \citenamefont {{Po}}, \citenamefont {{Vishwanath}},\
  and\ \citenamefont {{Kaxiras}}}]{CarrPoWannier}%
  \BibitemOpen
  \bibfield  {author} {\bibinfo {author} {\bibfnamefont {Stephen}\ \bibnamefont
  {{Carr}}}, \bibinfo {author} {\bibfnamefont {Shiang}\ \bibnamefont {{Fang}}},
  \bibinfo {author} {\bibfnamefont {Hoi~Chun}\ \bibnamefont {{Po}}}, \bibinfo
  {author} {\bibfnamefont {Ashvin}\ \bibnamefont {{Vishwanath}}}, \ and\
  \bibinfo {author} {\bibfnamefont {Efthimios}\ \bibnamefont {{Kaxiras}}},\
  }\bibfield  {title} {\enquote {\bibinfo {title} {{Derivation of Wannier
  orbitals and minimal-basis tight-binding Hamiltonians for twisted bilayer
  graphene: First-principles approach}},}\ }\href {\doibase
  10.1103/PhysRevResearch.1.033072} {\bibfield  {journal} {\bibinfo  {journal}
  {Physical Review Research}\ }\textbf {\bibinfo {volume} {1}},\ \bibinfo {eid}
  {033072} (\bibinfo {year} {2019})},\ \Eprint
  {http://arxiv.org/abs/1907.06282} {arXiv:1907.06282} \BibitemShut {NoStop}%
\bibitem [{\citenamefont {Po}\ \emph {et~al.}(2019)\citenamefont {Po},
  \citenamefont {Zou}, \citenamefont {Senthil},\ and\ \citenamefont
  {Vishwanath}}]{PoModels}%
  \BibitemOpen
  \bibfield  {author} {\bibinfo {author} {\bibfnamefont {Hoi~Chun}\
  \bibnamefont {Po}}, \bibinfo {author} {\bibfnamefont {Liujun}\ \bibnamefont
  {Zou}}, \bibinfo {author} {\bibfnamefont {T.}~\bibnamefont {Senthil}}, \ and\
  \bibinfo {author} {\bibfnamefont {Ashvin}\ \bibnamefont {Vishwanath}},\
  }\bibfield  {title} {\enquote {\bibinfo {title} {{Faithful tight-binding
  models and fragile topology of magic-angle bilayer graphene}},}\ }\href
  {\doibase 10.1103/PhysRevB.99.195455} {\bibfield  {journal} {\bibinfo
  {journal} {Phys. Rev. B}\ }\textbf {\bibinfo {volume} {99}},\ \bibinfo
  {pages} {195455} (\bibinfo {year} {2019})},\ \Eprint
  {http://arxiv.org/abs/1808.02482} {arXiv:1808.02482} \BibitemShut {NoStop}%
\bibitem [{\citenamefont {{Koshino}}\ \emph {et~al.}(2018)\citenamefont
  {{Koshino}}, \citenamefont {{Yuan}}, \citenamefont {{Koretsune}},
  \citenamefont {{Ochi}}, \citenamefont {{Kuroki}},\ and\ \citenamefont
  {{Fu}}}]{FuWannier}%
  \BibitemOpen
  \bibfield  {author} {\bibinfo {author} {\bibfnamefont {Mikito}\ \bibnamefont
  {{Koshino}}}, \bibinfo {author} {\bibfnamefont {Noah F.~Q.}\ \bibnamefont
  {{Yuan}}}, \bibinfo {author} {\bibfnamefont {Takashi}\ \bibnamefont
  {{Koretsune}}}, \bibinfo {author} {\bibfnamefont {Masayuki}\ \bibnamefont
  {{Ochi}}}, \bibinfo {author} {\bibfnamefont {Kazuhiko}\ \bibnamefont
  {{Kuroki}}}, \ and\ \bibinfo {author} {\bibfnamefont {Liang}\ \bibnamefont
  {{Fu}}},\ }\bibfield  {title} {\enquote {\bibinfo {title} {{Maximally
  Localized Wannier Orbitals and the Extended Hubbard Model for Twisted Bilayer
  Graphene}},}\ }\href {\doibase 10.1103/PhysRevX.8.031087} {\bibfield
  {journal} {\bibinfo  {journal} {Physical Review X}\ }\textbf {\bibinfo
  {volume} {8}},\ \bibinfo {eid} {031087} (\bibinfo {year} {2018})},\ \Eprint
  {http://arxiv.org/abs/1805.06819} {arXiv:1805.06819} \BibitemShut {NoStop}%
\bibitem [{\citenamefont {{Kang}}\ and\ \citenamefont
  {{Vafek}}(2018)}]{KangTBG}%
  \BibitemOpen
  \bibfield  {author} {\bibinfo {author} {\bibfnamefont {Jian}\ \bibnamefont
  {{Kang}}}\ and\ \bibinfo {author} {\bibfnamefont {Oskar}\ \bibnamefont
  {{Vafek}}},\ }\bibfield  {title} {\enquote {\bibinfo {title} {{Symmetry,
  Maximally Localized Wannier States, and a Low-Energy Model for Twisted
  Bilayer Graphene Narrow Bands}},}\ }\href {\doibase
  10.1103/PhysRevX.8.031088} {\bibfield  {journal} {\bibinfo  {journal}
  {Physical Review X}\ }\textbf {\bibinfo {volume} {8}},\ \bibinfo {eid}
  {031088} (\bibinfo {year} {2018})},\ \Eprint
  {http://arxiv.org/abs/1805.04918} {arXiv:1805.04918} \BibitemShut {NoStop}%
\bibitem [{\citenamefont {{Tarnopolsky}}\ \emph {et~al.}(2019)\citenamefont
  {{Tarnopolsky}}, \citenamefont {{Kruchkov}},\ and\ \citenamefont
  {{Vishwanath}}}]{TarnopolskyMagic}%
  \BibitemOpen
  \bibfield  {author} {\bibinfo {author} {\bibfnamefont {Grigory}\ \bibnamefont
  {{Tarnopolsky}}}, \bibinfo {author} {\bibfnamefont {Alex~Jura}\ \bibnamefont
  {{Kruchkov}}}, \ and\ \bibinfo {author} {\bibfnamefont {Ashvin}\ \bibnamefont
  {{Vishwanath}}},\ }\bibfield  {title} {\enquote {\bibinfo {title} {{Origin of
  Magic Angles in Twisted Bilayer Graphene}},}\ }\href {\doibase
  10.1103/PhysRevLett.122.106405} {\bibfield  {journal} {\bibinfo  {journal}
  {\prl}\ }\textbf {\bibinfo {volume} {122}},\ \bibinfo {eid} {106405}
  (\bibinfo {year} {2019})},\ \Eprint {http://arxiv.org/abs/1808.05250}
  {arXiv:1808.05250} \BibitemShut {NoStop}%
\bibitem [{\citenamefont {{Wu}}\ \emph
  {et~al.}(2019{\natexlab{a}})\citenamefont {{Wu}}, \citenamefont {{Jian}},\
  and\ \citenamefont {{Xu}}}]{XuWire}%
  \BibitemOpen
  \bibfield  {author} {\bibinfo {author} {\bibfnamefont {Xiao-Chuan}\
  \bibnamefont {{Wu}}}, \bibinfo {author} {\bibfnamefont {Chao-Ming}\
  \bibnamefont {{Jian}}}, \ and\ \bibinfo {author} {\bibfnamefont {Cenke}\
  \bibnamefont {{Xu}}},\ }\bibfield  {title} {\enquote {\bibinfo {title}
  {{Coupled-wire description of the correlated physics in twisted bilayer
  graphene}},}\ }\href {\doibase 10.1103/PhysRevB.99.161405} {\bibfield
  {journal} {\bibinfo  {journal} {\prb}\ }\textbf {\bibinfo {volume} {99}},\
  \bibinfo {eid} {161405} (\bibinfo {year} {2019}{\natexlab{a}})},\ \Eprint
  {http://arxiv.org/abs/1811.08442} {arXiv:1811.08442} \BibitemShut {NoStop}%
\bibitem [{\citenamefont {Manzeli}\ \emph {et~al.}(2017)\citenamefont
  {Manzeli}, \citenamefont {Ovchinnikov}, \citenamefont {Pasquier},
  \citenamefont {Yazyev},\ and\ \citenamefont {Kis}}]{TMDRev}%
  \BibitemOpen
  \bibfield  {author} {\bibinfo {author} {\bibfnamefont {Sajedeh}\ \bibnamefont
  {Manzeli}}, \bibinfo {author} {\bibfnamefont {Dmitry}\ \bibnamefont
  {Ovchinnikov}}, \bibinfo {author} {\bibfnamefont {Diego}\ \bibnamefont
  {Pasquier}}, \bibinfo {author} {\bibfnamefont {Oleg~V.}\ \bibnamefont
  {Yazyev}}, \ and\ \bibinfo {author} {\bibfnamefont {Andras}\ \bibnamefont
  {Kis}},\ }\bibfield  {title} {\enquote {\bibinfo {title} {2d transition metal
  dichalcogenides},}\ }\href {\doibase 10.1038/natrevmats.2017.33} {\bibfield
  {journal} {\bibinfo  {journal} {Nature Reviews Materials}\ }\textbf {\bibinfo
  {volume} {2}},\ \bibinfo {pages} {1–15} (\bibinfo {year}
  {2017})}\BibitemShut {NoStop}%
\bibitem [{\citenamefont {{Wang}}\ \emph
  {et~al.}(2020{\natexlab{a}})\citenamefont {{Wang}}, \citenamefont {{Shih}},
  \citenamefont {{Ghiotto}}, \citenamefont {{Xian}}, \citenamefont {{Rhodes}},
  \citenamefont {{Tan}}, \citenamefont {{Claassen}}, \citenamefont {{Kennes}},
  \citenamefont {{Bai}}, \citenamefont {{Kim}}, \citenamefont {{Watanabe}},
  \citenamefont {{Taniguchi}}, \citenamefont {{Zhu}}, \citenamefont {{Hone}},
  \citenamefont {{Rubio}}, \citenamefont {{Pasupathy}},\ and\ \citenamefont
  {{Dean}}}]{DeanTMD}%
  \BibitemOpen
  \bibfield  {author} {\bibinfo {author} {\bibfnamefont {Lei}\ \bibnamefont
  {{Wang}}}, \bibinfo {author} {\bibfnamefont {En-Min}\ \bibnamefont {{Shih}}},
  \bibinfo {author} {\bibfnamefont {Augusto}\ \bibnamefont {{Ghiotto}}},
  \bibinfo {author} {\bibfnamefont {Lede}\ \bibnamefont {{Xian}}}, \bibinfo
  {author} {\bibfnamefont {Daniel~A.}\ \bibnamefont {{Rhodes}}}, \bibinfo
  {author} {\bibfnamefont {Cheng}\ \bibnamefont {{Tan}}}, \bibinfo {author}
  {\bibfnamefont {Martin}\ \bibnamefont {{Claassen}}}, \bibinfo {author}
  {\bibfnamefont {Dante~M.}\ \bibnamefont {{Kennes}}}, \bibinfo {author}
  {\bibfnamefont {Yusong}\ \bibnamefont {{Bai}}}, \bibinfo {author}
  {\bibfnamefont {Bumho}\ \bibnamefont {{Kim}}}, \bibinfo {author}
  {\bibfnamefont {Kenji}\ \bibnamefont {{Watanabe}}}, \bibinfo {author}
  {\bibfnamefont {Takashi}\ \bibnamefont {{Taniguchi}}}, \bibinfo {author}
  {\bibfnamefont {Xiaoyang}\ \bibnamefont {{Zhu}}}, \bibinfo {author}
  {\bibfnamefont {James}\ \bibnamefont {{Hone}}}, \bibinfo {author}
  {\bibfnamefont {Angel}\ \bibnamefont {{Rubio}}}, \bibinfo {author}
  {\bibfnamefont {Abhay}\ \bibnamefont {{Pasupathy}}}, \ and\ \bibinfo {author}
  {\bibfnamefont {Cory~R.}\ \bibnamefont {{Dean}}},\ }\bibfield  {title}
  {\enquote {\bibinfo {title} {Correlated electronic phases in twisted bilayer
  transition metal dichalcogenides},}\ }\href {\doibase
  10.1038/s41563-020-0708-6} {\bibfield  {journal} {\bibinfo  {journal} {Nature
  Materials}\ }\textbf {\bibinfo {volume} {19}},\ \bibinfo {pages} {861–866}
  (\bibinfo {year} {2020}{\natexlab{a}})},\ \Eprint
  {http://arxiv.org/abs/1910.12147} {arXiv:1910.12147} \BibitemShut {NoStop}%
\bibitem [{\citenamefont {{Shi}}\ \emph {et~al.}(2020)\citenamefont {{Shi}},
  \citenamefont {{Shih}}, \citenamefont {{Gustafsson}}, \citenamefont
  {{Rhodes}}, \citenamefont {{Kim}}, \citenamefont {{Watanabe}}, \citenamefont
  {{Taniguchi}}, \citenamefont {{Papi{\'c}}}, \citenamefont {{Hone}},\ and\
  \citenamefont {{Dean}}}]{DeanHallTMD}%
  \BibitemOpen
  \bibfield  {author} {\bibinfo {author} {\bibfnamefont {Qianhui}\ \bibnamefont
  {{Shi}}}, \bibinfo {author} {\bibfnamefont {En-Min}\ \bibnamefont {{Shih}}},
  \bibinfo {author} {\bibfnamefont {Martin~V.}\ \bibnamefont {{Gustafsson}}},
  \bibinfo {author} {\bibfnamefont {Daniel~A.}\ \bibnamefont {{Rhodes}}},
  \bibinfo {author} {\bibfnamefont {Bumho}\ \bibnamefont {{Kim}}}, \bibinfo
  {author} {\bibfnamefont {Kenji}\ \bibnamefont {{Watanabe}}}, \bibinfo
  {author} {\bibfnamefont {Takashi}\ \bibnamefont {{Taniguchi}}}, \bibinfo
  {author} {\bibfnamefont {Zlatko}\ \bibnamefont {{Papi{\'c}}}}, \bibinfo
  {author} {\bibfnamefont {James}\ \bibnamefont {{Hone}}}, \ and\ \bibinfo
  {author} {\bibfnamefont {Cory~R.}\ \bibnamefont {{Dean}}},\ }\bibfield
  {title} {\enquote {\bibinfo {title} {{Odd- and even-denominator fractional
  quantum Hall states in monolayer WSe$_{2}$}},}\ }\href {\doibase
  10.1038/s41565-020-0685-6} {\bibfield  {journal} {\bibinfo  {journal} {Nature
  Nanotechnology}\ }\textbf {\bibinfo {volume} {15}},\ \bibinfo {pages}
  {569--573} (\bibinfo {year} {2020})},\ \Eprint
  {http://arxiv.org/abs/1911.04428} {arXiv:1911.04428} \BibitemShut {NoStop}%
\bibitem [{\citenamefont {{Wang}}\ \emph
  {et~al.}(2020{\natexlab{b}})\citenamefont {{Wang}}, \citenamefont {{Shi}},
  \citenamefont {{Shih}}, \citenamefont {{Zhou}}, \citenamefont {{Wu}},
  \citenamefont {{Bai}}, \citenamefont {{Rhodes}}, \citenamefont {{Barmak}},
  \citenamefont {{Hone}}, \citenamefont {{Dean}},\ and\ \citenamefont
  {{Zhu}}}]{DeanExciton}%
  \BibitemOpen
  \bibfield  {author} {\bibinfo {author} {\bibfnamefont {Jue}\ \bibnamefont
  {{Wang}}}, \bibinfo {author} {\bibfnamefont {Qianhui}\ \bibnamefont {{Shi}}},
  \bibinfo {author} {\bibfnamefont {En-Min}\ \bibnamefont {{Shih}}}, \bibinfo
  {author} {\bibfnamefont {Lin}\ \bibnamefont {{Zhou}}}, \bibinfo {author}
  {\bibfnamefont {Wenjing}\ \bibnamefont {{Wu}}}, \bibinfo {author}
  {\bibfnamefont {Yusong}\ \bibnamefont {{Bai}}}, \bibinfo {author}
  {\bibfnamefont {Daniel~A.}\ \bibnamefont {{Rhodes}}}, \bibinfo {author}
  {\bibfnamefont {Katayun}\ \bibnamefont {{Barmak}}}, \bibinfo {author}
  {\bibfnamefont {James}\ \bibnamefont {{Hone}}}, \bibinfo {author}
  {\bibfnamefont {Cory~R.}\ \bibnamefont {{Dean}}}, \ and\ \bibinfo {author}
  {\bibfnamefont {X.~Y.}\ \bibnamefont {{Zhu}}},\ }\bibfield  {title} {\enquote
  {\bibinfo {title} {{Excitonic Phase Transitions in MoSe2/WSe2
  Heterobilayers}},}\ }\href@noop {} {\  (\bibinfo {year}
  {2020}{\natexlab{b}})},\ \Eprint {http://arxiv.org/abs/2001.03812}
  {arXiv:2001.03812} \BibitemShut {NoStop}%
\bibitem [{\citenamefont {{Jin}}\ \emph {et~al.}(2019)\citenamefont {{Jin}},
  \citenamefont {{Regan}}, \citenamefont {{Yan}}, \citenamefont {{Iqbal Bakti
  Utama}}, \citenamefont {{Wang}}, \citenamefont {{Zhao}}, \citenamefont
  {{Qin}}, \citenamefont {{Yang}}, \citenamefont {{Zheng}}, \citenamefont
  {{Shi}}, \citenamefont {{Watanabe}}, \citenamefont {{Taniguchi}},
  \citenamefont {{Tongay}}, \citenamefont {{Zettl}},\ and\ \citenamefont
  {{Wang}}}]{FengExciton}%
  \BibitemOpen
  \bibfield  {author} {\bibinfo {author} {\bibfnamefont {Chenhao}\ \bibnamefont
  {{Jin}}}, \bibinfo {author} {\bibfnamefont {Emma~C.}\ \bibnamefont
  {{Regan}}}, \bibinfo {author} {\bibfnamefont {Aiming}\ \bibnamefont {{Yan}}},
  \bibinfo {author} {\bibfnamefont {M.}~\bibnamefont {{Iqbal Bakti Utama}}},
  \bibinfo {author} {\bibfnamefont {Danqing}\ \bibnamefont {{Wang}}}, \bibinfo
  {author} {\bibfnamefont {Sihan}\ \bibnamefont {{Zhao}}}, \bibinfo {author}
  {\bibfnamefont {Ying}\ \bibnamefont {{Qin}}}, \bibinfo {author}
  {\bibfnamefont {Sijie}\ \bibnamefont {{Yang}}}, \bibinfo {author}
  {\bibfnamefont {Zhiren}\ \bibnamefont {{Zheng}}}, \bibinfo {author}
  {\bibfnamefont {Shenyang}\ \bibnamefont {{Shi}}}, \bibinfo {author}
  {\bibfnamefont {Kenji}\ \bibnamefont {{Watanabe}}}, \bibinfo {author}
  {\bibfnamefont {Takashi}\ \bibnamefont {{Taniguchi}}}, \bibinfo {author}
  {\bibfnamefont {Sefaattin}\ \bibnamefont {{Tongay}}}, \bibinfo {author}
  {\bibfnamefont {Alex}\ \bibnamefont {{Zettl}}}, \ and\ \bibinfo {author}
  {\bibfnamefont {Feng}\ \bibnamefont {{Wang}}},\ }\bibfield  {title} {\enquote
  {\bibinfo {title} {{Observation of moir{\'e} excitons in WSe$_{2}$/WS$_{2}$
  heterostructure superlattices}},}\ }\href {\doibase
  10.1038/s41586-019-0976-y} {\bibfield  {journal} {\bibinfo  {journal} {\nat}\
  }\textbf {\bibinfo {volume} {567}},\ \bibinfo {pages} {76--80} (\bibinfo
  {year} {2019})},\ \Eprint {http://arxiv.org/abs/1812.09815}
  {arXiv:1812.09815} \BibitemShut {NoStop}%
\bibitem [{\citenamefont {Shimazaki}\ \emph {et~al.}(2020)\citenamefont
  {Shimazaki}, \citenamefont {Schwartz}, \citenamefont {Watanabe},
  \citenamefont {Taniguchi}, \citenamefont {Kroner},\ and\ \citenamefont
  {{Imamo{\u{g}}lu}}}]{Imamoglu2019}%
  \BibitemOpen
  \bibfield  {author} {\bibinfo {author} {\bibfnamefont {Yuya}\ \bibnamefont
  {Shimazaki}}, \bibinfo {author} {\bibfnamefont {Ido}\ \bibnamefont
  {Schwartz}}, \bibinfo {author} {\bibfnamefont {Kenji}\ \bibnamefont
  {Watanabe}}, \bibinfo {author} {\bibfnamefont {Takashi}\ \bibnamefont
  {Taniguchi}}, \bibinfo {author} {\bibfnamefont {Martin}\ \bibnamefont
  {Kroner}}, \ and\ \bibinfo {author} {\bibfnamefont {Ata{\c{c}}}\ \bibnamefont
  {{Imamo{\u{g}}lu}}},\ }\bibfield  {title} {\enquote {\bibinfo {title}
  {Strongly correlated electrons and hybrid excitons in a moiré
  heterostructure},}\ }\href {\doibase 10.1038/s41586-020-2191-2} {\bibfield
  {journal} {\bibinfo  {journal} {Nature}\ }\textbf {\bibinfo {volume} {580}},\
  \bibinfo {pages} {472–477} (\bibinfo {year} {2020})},\ \Eprint
  {http://arxiv.org/abs/1910.13322} {arXiv:1910.13322} \BibitemShut {NoStop}%
\bibitem [{\citenamefont {{Sung}}\ \emph {et~al.}(2020)\citenamefont {{Sung}},
  \citenamefont {{Zhou}}, \citenamefont {{Scuri}}, \citenamefont
  {{Z{\'o}lyomi}}, \citenamefont {{Andersen}}, \citenamefont {{Yoo}},
  \citenamefont {{Wild}}, \citenamefont {{Joe}}, \citenamefont {{Gelly}},
  \citenamefont {{Heo}}, \citenamefont {{Magorrian}}, \citenamefont
  {{B{\'e}rub{\'e}}}, \citenamefont {{Valdivia}}, \citenamefont {{Taniguchi}},
  \citenamefont {{Watanabe}}, \citenamefont {{Lukin}}, \citenamefont {{Kim}},
  \citenamefont {{Fal'ko}},\ and\ \citenamefont {{Park}}}]{Park2020}%
  \BibitemOpen
  \bibfield  {author} {\bibinfo {author} {\bibfnamefont {Jiho}\ \bibnamefont
  {{Sung}}}, \bibinfo {author} {\bibfnamefont {You}\ \bibnamefont {{Zhou}}},
  \bibinfo {author} {\bibfnamefont {Giovanni}\ \bibnamefont {{Scuri}}},
  \bibinfo {author} {\bibfnamefont {Viktor}\ \bibnamefont {{Z{\'o}lyomi}}},
  \bibinfo {author} {\bibfnamefont {Trond~I.}\ \bibnamefont {{Andersen}}},
  \bibinfo {author} {\bibfnamefont {Hyobin}\ \bibnamefont {{Yoo}}}, \bibinfo
  {author} {\bibfnamefont {Dominik~S.}\ \bibnamefont {{Wild}}}, \bibinfo
  {author} {\bibfnamefont {Andrew~Y.}\ \bibnamefont {{Joe}}}, \bibinfo {author}
  {\bibfnamefont {Ryan~J.}\ \bibnamefont {{Gelly}}}, \bibinfo {author}
  {\bibfnamefont {Hoseok}\ \bibnamefont {{Heo}}}, \bibinfo {author}
  {\bibfnamefont {Samuel~J.}\ \bibnamefont {{Magorrian}}}, \bibinfo {author}
  {\bibfnamefont {Damien}\ \bibnamefont {{B{\'e}rub{\'e}}}}, \bibinfo {author}
  {\bibfnamefont {Andr{\'e}s M.~Mier}\ \bibnamefont {{Valdivia}}}, \bibinfo
  {author} {\bibfnamefont {Takashi}\ \bibnamefont {{Taniguchi}}}, \bibinfo
  {author} {\bibfnamefont {Kenji}\ \bibnamefont {{Watanabe}}}, \bibinfo
  {author} {\bibfnamefont {Mikhail~D.}\ \bibnamefont {{Lukin}}}, \bibinfo
  {author} {\bibfnamefont {Philip}\ \bibnamefont {{Kim}}}, \bibinfo {author}
  {\bibfnamefont {Vladimir~I.}\ \bibnamefont {{Fal'ko}}}, \ and\ \bibinfo
  {author} {\bibfnamefont {Hongkun}\ \bibnamefont {{Park}}},\ }\bibfield
  {title} {\enquote {\bibinfo {title} {{Broken mirror symmetry in excitonic
  response of reconstructed domains in twisted MoSe$_{2}$/MoSe$_{2}$
  bilayers}},}\ }\href {\doibase 10.1038/s41565-020-0728-z} {\bibfield
  {journal} {\bibinfo  {journal} {Nature Nanotechnology}\ }\textbf {\bibinfo
  {volume} {15}},\ \bibinfo {pages} {750--754} (\bibinfo {year} {2020})},\
  \Eprint {http://arxiv.org/abs/2001.01157} {arXiv:2001.01157} \BibitemShut
  {NoStop}%
\bibitem [{\citenamefont {{Bi}}\ and\ \citenamefont
  {{Fu}}(2019)}]{BiFuExcitonicTMD}%
  \BibitemOpen
  \bibfield  {author} {\bibinfo {author} {\bibfnamefont {Zhen}\ \bibnamefont
  {{Bi}}}\ and\ \bibinfo {author} {\bibfnamefont {Liang}\ \bibnamefont
  {{Fu}}},\ }\bibfield  {title} {\enquote {\bibinfo {title} {{Excitonic density
  wave and spin-valley superfluid in bilayer transition metal
  dichalcogenide}},}\ }\href@noop {} {\  (\bibinfo {year} {2019})},\ \Eprint
  {http://arxiv.org/abs/1911.04493} {arXiv:1911.04493} \BibitemShut {NoStop}%
\bibitem [{\citenamefont {{Wu}}\ \emph {et~al.}(2018)\citenamefont {{Wu}},
  \citenamefont {{Lovorn}}, \citenamefont {{Tutuc}},\ and\ \citenamefont
  {{MacDonald}}}]{MacDonaldTMD}%
  \BibitemOpen
  \bibfield  {author} {\bibinfo {author} {\bibfnamefont {Fengcheng}\
  \bibnamefont {{Wu}}}, \bibinfo {author} {\bibfnamefont {Timothy}\
  \bibnamefont {{Lovorn}}}, \bibinfo {author} {\bibfnamefont {Emanuel}\
  \bibnamefont {{Tutuc}}}, \ and\ \bibinfo {author} {\bibfnamefont {A.~H.}\
  \bibnamefont {{MacDonald}}},\ }\bibfield  {title} {\enquote {\bibinfo {title}
  {{Hubbard Model Physics in Transition Metal Dichalcogenide Moir{\'e}
  Bands}},}\ }\href {\doibase 10.1103/PhysRevLett.121.026402} {\bibfield
  {journal} {\bibinfo  {journal} {\prl}\ }\textbf {\bibinfo {volume} {121}},\
  \bibinfo {eid} {026402} (\bibinfo {year} {2018})},\ \Eprint
  {http://arxiv.org/abs/1804.03151} {arXiv:1804.03151} \BibitemShut {NoStop}%
\bibitem [{\citenamefont {{Wu}}\ \emph
  {et~al.}(2019{\natexlab{b}})\citenamefont {{Wu}}, \citenamefont {{Lovorn}},
  \citenamefont {{Tutuc}}, \citenamefont {{Martin}},\ and\ \citenamefont
  {{MacDonald}}}]{MacDonaldTopoTMD}%
  \BibitemOpen
  \bibfield  {author} {\bibinfo {author} {\bibfnamefont {Fengcheng}\
  \bibnamefont {{Wu}}}, \bibinfo {author} {\bibfnamefont {Timothy}\
  \bibnamefont {{Lovorn}}}, \bibinfo {author} {\bibfnamefont {Emanuel}\
  \bibnamefont {{Tutuc}}}, \bibinfo {author} {\bibfnamefont {Ivar}\
  \bibnamefont {{Martin}}}, \ and\ \bibinfo {author} {\bibfnamefont {A.~H.}\
  \bibnamefont {{MacDonald}}},\ }\bibfield  {title} {\enquote {\bibinfo {title}
  {{Topological Insulators in Twisted Transition Metal Dichalcogenide
  Homobilayers}},}\ }\href {\doibase 10.1103/PhysRevLett.122.086402} {\bibfield
   {journal} {\bibinfo  {journal} {\prl}\ }\textbf {\bibinfo {volume} {122}},\
  \bibinfo {eid} {086402} (\bibinfo {year} {2019}{\natexlab{b}})},\ \Eprint
  {http://arxiv.org/abs/1807.03311} {arXiv:1807.03311} \BibitemShut {NoStop}%
\bibitem [{\citenamefont {{Xian}}\ \emph {et~al.}(2020)\citenamefont {{Xian}},
  \citenamefont {{Claassen}}, \citenamefont {{Kiese}}, \citenamefont
  {{Scherer}}, \citenamefont {{Trebst}}, \citenamefont {{Kennes}},\ and\
  \citenamefont {{Rubio}}}]{TMDRealization}%
  \BibitemOpen
  \bibfield  {author} {\bibinfo {author} {\bibfnamefont {Lede}\ \bibnamefont
  {{Xian}}}, \bibinfo {author} {\bibfnamefont {Martin}\ \bibnamefont
  {{Claassen}}}, \bibinfo {author} {\bibfnamefont {Dominik}\ \bibnamefont
  {{Kiese}}}, \bibinfo {author} {\bibfnamefont {Michael~M.}\ \bibnamefont
  {{Scherer}}}, \bibinfo {author} {\bibfnamefont {Simon}\ \bibnamefont
  {{Trebst}}}, \bibinfo {author} {\bibfnamefont {Dante~M.}\ \bibnamefont
  {{Kennes}}}, \ and\ \bibinfo {author} {\bibfnamefont {Angel}\ \bibnamefont
  {{Rubio}}},\ }\bibfield  {title} {\enquote {\bibinfo {title} {{Realization of
  Nearly Dispersionless Bands with Strong Orbital Anisotropy from Destructive
  Interference in Twisted Bilayer MoS2}},}\ }\href@noop {} {\  (\bibinfo {year}
  {2020})},\ \Eprint {http://arxiv.org/abs/2004.02964} {arXiv:2004.02964}
  \BibitemShut {NoStop}%
\bibitem [{\citenamefont {{Tang}}\ \emph {et~al.}(2020)\citenamefont {{Tang}},
  \citenamefont {{Li}}, \citenamefont {{Li}}, \citenamefont {{Xu}},
  \citenamefont {{Liu}}, \citenamefont {{Barmak}}, \citenamefont {{Watanabe}},
  \citenamefont {{Taniguchi}}, \citenamefont {{MacDonald}}, \citenamefont
  {{Shan}},\ and\ \citenamefont {{Mak}}}]{FaiTMD}%
  \BibitemOpen
  \bibfield  {author} {\bibinfo {author} {\bibfnamefont {Yanhao}\ \bibnamefont
  {{Tang}}}, \bibinfo {author} {\bibfnamefont {Lizhong}\ \bibnamefont {{Li}}},
  \bibinfo {author} {\bibfnamefont {Tingxin}\ \bibnamefont {{Li}}}, \bibinfo
  {author} {\bibfnamefont {Yang}\ \bibnamefont {{Xu}}}, \bibinfo {author}
  {\bibfnamefont {Song}\ \bibnamefont {{Liu}}}, \bibinfo {author}
  {\bibfnamefont {Katayun}\ \bibnamefont {{Barmak}}}, \bibinfo {author}
  {\bibfnamefont {Kenji}\ \bibnamefont {{Watanabe}}}, \bibinfo {author}
  {\bibfnamefont {Takashi}\ \bibnamefont {{Taniguchi}}}, \bibinfo {author}
  {\bibfnamefont {Allan~H}\ \bibnamefont {{MacDonald}}}, \bibinfo {author}
  {\bibfnamefont {Jie}\ \bibnamefont {{Shan}}}, \ and\ \bibinfo {author}
  {\bibfnamefont {Kin~Fai}\ \bibnamefont {{Mak}}},\ }\bibfield  {title}
  {\enquote {\bibinfo {title} {{Simulation of Hubbard model physics in
  WSe$_2$/WS$_2$ moiré superlattices}},}\ }\href {\doibase
  10.1038/s41586-020-2085-3} {\bibfield  {journal} {\bibinfo  {journal}
  {Nature}\ }\textbf {\bibinfo {volume} {579}},\ \bibinfo {pages} {353–358}
  (\bibinfo {year} {2020})},\ \Eprint {http://arxiv.org/abs/1910.08673}
  {arXiv:1910.08673} \BibitemShut {NoStop}%
\bibitem [{\citenamefont {{Regan}}\ \emph {et~al.}(2020)\citenamefont
  {{Regan}}, \citenamefont {{Wang}}, \citenamefont {{Jin}}, \citenamefont
  {{Bakti Utama}}, \citenamefont {{Gao}}, \citenamefont {{Wei}}, \citenamefont
  {{Zhao}}, \citenamefont {{Zhao}}, \citenamefont {{Yumigeta}}, \citenamefont
  {{Blei}}, \citenamefont {{Carlstroem}}, \citenamefont {{Watanabe}},
  \citenamefont {{Taniguchi}}, \citenamefont {{Tongay}}, \citenamefont
  {{Crommie}}, \citenamefont {{Zettl}},\ and\ \citenamefont
  {{Wang}}}]{BerkeleyTMD}%
  \BibitemOpen
  \bibfield  {author} {\bibinfo {author} {\bibfnamefont {Emma~C.}\ \bibnamefont
  {{Regan}}}, \bibinfo {author} {\bibfnamefont {Danqing}\ \bibnamefont
  {{Wang}}}, \bibinfo {author} {\bibfnamefont {Chenhao}\ \bibnamefont {{Jin}}},
  \bibinfo {author} {\bibfnamefont {M.~Iqbal}\ \bibnamefont {{Bakti Utama}}},
  \bibinfo {author} {\bibfnamefont {Beini}\ \bibnamefont {{Gao}}}, \bibinfo
  {author} {\bibfnamefont {Xin}\ \bibnamefont {{Wei}}}, \bibinfo {author}
  {\bibfnamefont {Sihan}\ \bibnamefont {{Zhao}}}, \bibinfo {author}
  {\bibfnamefont {Wenyu}\ \bibnamefont {{Zhao}}}, \bibinfo {author}
  {\bibfnamefont {Kentaro}\ \bibnamefont {{Yumigeta}}}, \bibinfo {author}
  {\bibfnamefont {Mark}\ \bibnamefont {{Blei}}}, \bibinfo {author}
  {\bibfnamefont {Johan}\ \bibnamefont {{Carlstroem}}}, \bibinfo {author}
  {\bibfnamefont {Kenji}\ \bibnamefont {{Watanabe}}}, \bibinfo {author}
  {\bibfnamefont {Takashi}\ \bibnamefont {{Taniguchi}}}, \bibinfo {author}
  {\bibfnamefont {Sefaattin}\ \bibnamefont {{Tongay}}}, \bibinfo {author}
  {\bibfnamefont {Michael}\ \bibnamefont {{Crommie}}}, \bibinfo {author}
  {\bibfnamefont {Alex}\ \bibnamefont {{Zettl}}}, \ and\ \bibinfo {author}
  {\bibfnamefont {Feng}\ \bibnamefont {{Wang}}},\ }\bibfield  {title} {\enquote
  {\bibinfo {title} {{Mott and generalized Wigner crystal states in
  WSe$_2$/WS$_2$ moiré superlattices}},}\ }\href {\doibase
  10.1038/s41586-020-2092-4} {\bibfield  {journal} {\bibinfo  {journal}
  {Nature}\ }\textbf {\bibinfo {volume} {579}},\ \bibinfo {pages} {359–363}
  (\bibinfo {year} {2020})},\ \Eprint {http://arxiv.org/abs/1910.09047}
  {arXiv:1910.09047} \BibitemShut {NoStop}%
\bibitem [{\citenamefont {{Xu}}\ \emph {et~al.}(2020)\citenamefont {{Xu}},
  \citenamefont {{Liu}}, \citenamefont {{Rhodes}}, \citenamefont {{Watanabe}},
  \citenamefont {{Taniguchi}}, \citenamefont {{Hone}}, \citenamefont {{Elser}},
  \citenamefont {{Mak}},\ and\ \citenamefont {{Shan}}}]{ShanTMD}%
  \BibitemOpen
  \bibfield  {author} {\bibinfo {author} {\bibfnamefont {Yang}\ \bibnamefont
  {{Xu}}}, \bibinfo {author} {\bibfnamefont {Song}\ \bibnamefont {{Liu}}},
  \bibinfo {author} {\bibfnamefont {Daniel~A}\ \bibnamefont {{Rhodes}}},
  \bibinfo {author} {\bibfnamefont {Kenji}\ \bibnamefont {{Watanabe}}},
  \bibinfo {author} {\bibfnamefont {Takashi}\ \bibnamefont {{Taniguchi}}},
  \bibinfo {author} {\bibfnamefont {James}\ \bibnamefont {{Hone}}}, \bibinfo
  {author} {\bibfnamefont {Veit}\ \bibnamefont {{Elser}}}, \bibinfo {author}
  {\bibfnamefont {Kin~Fai}\ \bibnamefont {{Mak}}}, \ and\ \bibinfo {author}
  {\bibfnamefont {Jie}\ \bibnamefont {{Shan}}},\ }\bibfield  {title} {\enquote
  {\bibinfo {title} {{Abundance of correlated insulating states at fractional
  fillings of WSe$_{2}$/WS$_{2}$ moir{\'e} superlattices}},}\ }\href@noop {} {\
   (\bibinfo {year} {2020})},\ \Eprint {http://arxiv.org/abs/2007.11128}
  {arXiv:2007.11128} \BibitemShut {NoStop}%
\bibitem [{\citenamefont {{Zhang}}\ \emph {et~al.}(2019)\citenamefont
  {{Zhang}}, \citenamefont {{Yuan}},\ and\ \citenamefont
  {{Fu}}}]{FuChargeTransfer}%
  \BibitemOpen
  \bibfield  {author} {\bibinfo {author} {\bibfnamefont {Yang}\ \bibnamefont
  {{Zhang}}}, \bibinfo {author} {\bibfnamefont {Noah F.~Q.}\ \bibnamefont
  {{Yuan}}}, \ and\ \bibinfo {author} {\bibfnamefont {Liang}\ \bibnamefont
  {{Fu}}},\ }\bibfield  {title} {\enquote {\bibinfo {title} {{Moir{\'e} quantum
  chemistry: charge transfer in transition metal dichalcogenide
  superlattices}},}\ }\href@noop {} {\  (\bibinfo {year} {2019})},\ \Eprint
  {http://arxiv.org/abs/1910.14061} {arXiv:1910.14061} \BibitemShut {NoStop}%
\bibitem [{Efo()}]{Efoot:graphene}%
  \BibitemOpen
  \href@noop {} {}\bibinfo {note} {Localized and symmetric Wannier orbitals are
  not even possible in twisted bilayer graphene \cite{PoModels}}\BibitemShut
  {NoStop}%
\bibitem [{\citenamefont {Zaanen}\ \emph {et~al.}(1985)\citenamefont {Zaanen},
  \citenamefont {Sawatzky},\ and\ \citenamefont
  {Allen}}]{chargeTransferInsulator}%
  \BibitemOpen
  \bibfield  {author} {\bibinfo {author} {\bibfnamefont {J.}~\bibnamefont
  {Zaanen}}, \bibinfo {author} {\bibfnamefont {G.~A.}\ \bibnamefont
  {Sawatzky}}, \ and\ \bibinfo {author} {\bibfnamefont {J.~W.}\ \bibnamefont
  {Allen}},\ }\bibfield  {title} {\enquote {\bibinfo {title} {{Band gaps and
  electronic structure of transition-metal compounds}},}\ }\href {\doibase
  10.1103/PhysRevLett.55.418} {\bibfield  {journal} {\bibinfo  {journal} {Phys.
  Rev. Lett.}\ }\textbf {\bibinfo {volume} {55}},\ \bibinfo {pages} {418--421}
  (\bibinfo {year} {1985})}\BibitemShut {NoStop}%
\bibitem [{\citenamefont {Emery}(1987)}]{EmeryHighTc}%
  \BibitemOpen
  \bibfield  {author} {\bibinfo {author} {\bibfnamefont {V.~J.}\ \bibnamefont
  {Emery}},\ }\bibfield  {title} {\enquote {\bibinfo {title} {{Theory of
  high-${\mathrm{T}}_{\mathrm{c}}$ superconductivity in oxides}},}\ }\href
  {\doibase 10.1103/PhysRevLett.58.2794} {\bibfield  {journal} {\bibinfo
  {journal} {Phys. Rev. Lett.}\ }\textbf {\bibinfo {volume} {58}},\ \bibinfo
  {pages} {2794--2797} (\bibinfo {year} {1987})}\BibitemShut {NoStop}%
\bibitem [{\citenamefont {Zhang}\ and\ \citenamefont
  {Rice}(1988)}]{ZhangRiceCuO}%
  \BibitemOpen
  \bibfield  {author} {\bibinfo {author} {\bibfnamefont {F.~C.}\ \bibnamefont
  {Zhang}}\ and\ \bibinfo {author} {\bibfnamefont {T.~M.}\ \bibnamefont
  {Rice}},\ }\bibfield  {title} {\enquote {\bibinfo {title} {{Effective
  Hamiltonian for the superconducting Cu oxides}},}\ }\href {\doibase
  10.1103/PhysRevB.37.3759} {\bibfield  {journal} {\bibinfo  {journal} {Phys.
  Rev. B}\ }\textbf {\bibinfo {volume} {37}},\ \bibinfo {pages} {3759--3761}
  (\bibinfo {year} {1988})}\BibitemShut {NoStop}%
\bibitem [{\citenamefont {Bar-Yam}(1991)}]{BarYamTwoComponent}%
  \BibitemOpen
  \bibfield  {author} {\bibinfo {author} {\bibfnamefont {Y.}~\bibnamefont
  {Bar-Yam}},\ }\bibfield  {title} {\enquote {\bibinfo {title} {{Two-component
  superconductivity. I. Introduction and phenomenology}},}\ }\href {\doibase
  10.1103/PhysRevB.43.359} {\bibfield  {journal} {\bibinfo  {journal} {Phys.
  Rev. B}\ }\textbf {\bibinfo {volume} {43}},\ \bibinfo {pages} {359--377}
  (\bibinfo {year} {1991})}\BibitemShut {NoStop}%
\bibitem [{\citenamefont {Abrahams}\ \emph {et~al.}(1987)\citenamefont
  {Abrahams}, \citenamefont {Schmitt-Rink},\ and\ \citenamefont
  {Varma}}]{VarmaChargeTransfer}%
  \BibitemOpen
  \bibfield  {author} {\bibinfo {author} {\bibfnamefont {Elihu}\ \bibnamefont
  {Abrahams}}, \bibinfo {author} {\bibfnamefont {S.}~\bibnamefont
  {Schmitt-Rink}}, \ and\ \bibinfo {author} {\bibfnamefont {C.M.}\ \bibnamefont
  {Varma}},\ }\bibfield  {title} {\enquote {\bibinfo {title} {{High-Tc
  superconductivity mediated by charge transfer excitations}},}\ }\href
  {\doibase 10.1016/0378-4363(87)90204-X} {\bibfield  {journal} {\bibinfo
  {journal} {Physica B+C}\ }\textbf {\bibinfo {volume} {148}},\ \bibinfo
  {pages} {257 -- 259} (\bibinfo {year} {1987})}\BibitemShut {NoStop}%
\bibitem [{\citenamefont {Grilli}\ \emph {et~al.}(1991)\citenamefont {Grilli},
  \citenamefont {Raimondi}, \citenamefont {Castellani}, \citenamefont
  {Di~Castro},\ and\ \citenamefont {Kotliar}}]{KotliarChargeTransfer}%
  \BibitemOpen
  \bibfield  {author} {\bibinfo {author} {\bibfnamefont {M.}~\bibnamefont
  {Grilli}}, \bibinfo {author} {\bibfnamefont {R.}~\bibnamefont {Raimondi}},
  \bibinfo {author} {\bibfnamefont {C.}~\bibnamefont {Castellani}}, \bibinfo
  {author} {\bibfnamefont {C.}~\bibnamefont {Di~Castro}}, \ and\ \bibinfo
  {author} {\bibfnamefont {G.}~\bibnamefont {Kotliar}},\ }\bibfield  {title}
  {\enquote {\bibinfo {title} {{Superconductivity, phase separation, and
  charge-transfer instability in the U=\ensuremath{\infty} limit of the
  three-band model of the ${\mathrm{CuO}}_{2}$ planes}},}\ }\href {\doibase
  10.1103/PhysRevLett.67.259} {\bibfield  {journal} {\bibinfo  {journal} {Phys.
  Rev. Lett.}\ }\textbf {\bibinfo {volume} {67}},\ \bibinfo {pages} {259--262}
  (\bibinfo {year} {1991})}\BibitemShut {NoStop}%
\bibitem [{\citenamefont {{Gurarie}}\ and\ \citenamefont
  {{Radzihovsky}}(2007)}]{GurarieRadzihovskyFeshbach}%
  \BibitemOpen
  \bibfield  {author} {\bibinfo {author} {\bibfnamefont {V.}~\bibnamefont
  {{Gurarie}}}\ and\ \bibinfo {author} {\bibfnamefont {L.}~\bibnamefont
  {{Radzihovsky}}},\ }\bibfield  {title} {\enquote {\bibinfo {title}
  {{Resonantly paired fermionic superfluids}},}\ }\href {\doibase
  10.1016/j.aop.2006.10.009} {\bibfield  {journal} {\bibinfo  {journal} {Annals
  of Physics}\ }\textbf {\bibinfo {volume} {322}},\ \bibinfo {pages} {2--119}
  (\bibinfo {year} {2007})},\ \Eprint {http://arxiv.org/abs/cond-mat/0611022}
  {arXiv:cond-mat/0611022} \BibitemShut {NoStop}%
\bibitem [{\citenamefont {{Chen}}\ \emph {et~al.}(2005)\citenamefont {{Chen}},
  \citenamefont {{Stajic}}, \citenamefont {{Tan}},\ and\ \citenamefont
  {{Levin}}}]{BECBCS}%
  \BibitemOpen
  \bibfield  {author} {\bibinfo {author} {\bibfnamefont {Qijin}\ \bibnamefont
  {{Chen}}}, \bibinfo {author} {\bibfnamefont {Jelena}\ \bibnamefont
  {{Stajic}}}, \bibinfo {author} {\bibfnamefont {Shina}\ \bibnamefont {{Tan}}},
  \ and\ \bibinfo {author} {\bibfnamefont {K.}~\bibnamefont {{Levin}}},\
  }\bibfield  {title} {\enquote {\bibinfo {title} {{BCS BEC crossover: From
  high temperature superconductors to ultracold superfluids}},}\ }\href
  {\doibase 10.1016/j.physrep.2005.02.005} {\bibfield  {journal} {\bibinfo
  {journal} {Phys. Rep.}\ }\textbf {\bibinfo {volume} {412}},\ \bibinfo {pages}
  {1--88} (\bibinfo {year} {2005})},\ \Eprint
  {http://arxiv.org/abs/cond-mat/0404274} {arXiv:cond-mat/0404274} \BibitemShut
  {NoStop}%
\bibitem [{\citenamefont {Liu}\ and\ \citenamefont
  {Zhang}()}]{LiuZhandPrivate}%
  \BibitemOpen
  \bibfield  {author} {\bibinfo {author} {\bibfnamefont {Tongtong}\
  \bibnamefont {Liu}}\ and\ \bibinfo {author} {\bibfnamefont {Yang}\
  \bibnamefont {Zhang}},\ }\href@noop {} {}\bibinfo {howpublished} {private
  communication. Calculated using the screened Coulomb interaction shown in
  \eqnref{eq:2gate} and the resulting localized Wannier orbitals.}\BibitemShut
  {Stop}%
\bibitem [{Dfo()}]{Dfoot:Vac}%
  \BibitemOpen
  \href@noop {} {}\bibinfo {note} {$V_c = 3V_1 + 3V_3 + 6V_4 + \cdots$ and
  $V_a= 6V_2 + 6V_5 + \cdots$ where $V_n$ is the repulsion between
  $n^\text{th}$ nearest-neighbors, shown in \figref{fig:neighbors} of the
  Appendix.}\BibitemShut {Stop}%
\bibitem [{\citenamefont {White}\ \emph {et~al.}(1992)\citenamefont {White},
  \citenamefont {Chakravarty}, \citenamefont {Gelfand},\ and\ \citenamefont
  {Kivelson}}]{WhiteMolecules}%
  \BibitemOpen
  \bibfield  {author} {\bibinfo {author} {\bibfnamefont {Steven~R.}\
  \bibnamefont {White}}, \bibinfo {author} {\bibfnamefont {Sudip}\ \bibnamefont
  {Chakravarty}}, \bibinfo {author} {\bibfnamefont {Martin~P.}\ \bibnamefont
  {Gelfand}}, \ and\ \bibinfo {author} {\bibfnamefont {Steven~A.}\ \bibnamefont
  {Kivelson}},\ }\bibfield  {title} {\enquote {\bibinfo {title} {{Pair binding
  in small Hubbard-model molecules}},}\ }\href {\doibase
  10.1103/PhysRevB.45.5062} {\bibfield  {journal} {\bibinfo  {journal} {Phys.
  Rev. B}\ }\textbf {\bibinfo {volume} {45}},\ \bibinfo {pages} {5062--5065}
  (\bibinfo {year} {1992})}\BibitemShut {NoStop}%
\bibitem [{\citenamefont {{Yao}}\ \emph {et~al.}(2007)\citenamefont {{Yao}},
  \citenamefont {{Tsai}},\ and\ \citenamefont
  {{Kivelson}}}]{KivelsonCheckerboard}%
  \BibitemOpen
  \bibfield  {author} {\bibinfo {author} {\bibfnamefont {Hong}\ \bibnamefont
  {{Yao}}}, \bibinfo {author} {\bibfnamefont {Wei-Feng}\ \bibnamefont
  {{Tsai}}}, \ and\ \bibinfo {author} {\bibfnamefont {Steven~A.}\ \bibnamefont
  {{Kivelson}}},\ }\bibfield  {title} {\enquote {\bibinfo {title} {{Myriad
  phases of the checkerboard Hubbard model}},}\ }\href {\doibase
  10.1103/PhysRevB.76.161104} {\bibfield  {journal} {\bibinfo  {journal}
  {\prb}\ }\textbf {\bibinfo {volume} {76}},\ \bibinfo {eid} {161104} (\bibinfo
  {year} {2007})},\ \Eprint {http://arxiv.org/abs/0706.0761} {arXiv:0706.0761}
  \BibitemShut {NoStop}%
\bibitem [{\citenamefont {{Isaev}}\ \emph {et~al.}(2010)\citenamefont
  {{Isaev}}, \citenamefont {{Ortiz}},\ and\ \citenamefont
  {{Batista}}}]{IsaevOrtiz}%
  \BibitemOpen
  \bibfield  {author} {\bibinfo {author} {\bibfnamefont {L.}~\bibnamefont
  {{Isaev}}}, \bibinfo {author} {\bibfnamefont {G.}~\bibnamefont {{Ortiz}}}, \
  and\ \bibinfo {author} {\bibfnamefont {C.~D.}\ \bibnamefont {{Batista}}},\
  }\bibfield  {title} {\enquote {\bibinfo {title} {{Superconductivity in
  Strongly Repulsive Fermions: The Role of Kinetic-Energy Frustration}},}\
  }\href {\doibase 10.1103/PhysRevLett.105.187002} {\bibfield  {journal}
  {\bibinfo  {journal} {\prl}\ }\textbf {\bibinfo {volume} {105}},\ \bibinfo
  {eid} {187002} (\bibinfo {year} {2010})},\ \Eprint
  {http://arxiv.org/abs/1003.5207} {arXiv:1003.5207} \BibitemShut {NoStop}%
\bibitem [{\citenamefont {Slagle}\ and\ \citenamefont {Kim}(2019)}]{SlagleKim}%
  \BibitemOpen
  \bibfield  {author} {\bibinfo {author} {\bibfnamefont {Kevin}\ \bibnamefont
  {Slagle}}\ and\ \bibinfo {author} {\bibfnamefont {Yong~Baek}\ \bibnamefont
  {Kim}},\ }\bibfield  {title} {\enquote {\bibinfo {title} {{A Simple Mechanism
  for Unconventional Superconductivity in a Repulsive Fermion Model}},}\ }\href
  {\doibase 10.21468/SciPostPhys.6.2.016} {\bibfield  {journal} {\bibinfo
  {journal} {SciPost Phys.}\ }\textbf {\bibinfo {volume} {6}},\ \bibinfo
  {pages} {16} (\bibinfo {year} {2019})},\ \Eprint
  {http://arxiv.org/abs/1805.05331} {arXiv:1805.05331} \BibitemShut {NoStop}%
\bibitem [{\citenamefont {{Padhi}}\ \emph {et~al.}(2018)\citenamefont
  {{Padhi}}, \citenamefont {{Setty}},\ and\ \citenamefont
  {{Phillips}}}]{PhilipWigner}%
  \BibitemOpen
  \bibfield  {author} {\bibinfo {author} {\bibfnamefont {Bikash}\ \bibnamefont
  {{Padhi}}}, \bibinfo {author} {\bibfnamefont {Chandan}\ \bibnamefont
  {{Setty}}}, \ and\ \bibinfo {author} {\bibfnamefont {Philip~W.}\ \bibnamefont
  {{Phillips}}},\ }\bibfield  {title} {\enquote {\bibinfo {title} {{Doped
  Twisted Bilayer Graphene near Magic Angles: Proximity to Wigner
  Crystallization, Not Mott Insulation}},}\ }\href {\doibase
  10.1021/acs.nanolett.8b02033} {\bibfield  {journal} {\bibinfo  {journal}
  {Nano Letters}\ }\textbf {\bibinfo {volume} {18}},\ \bibinfo {pages}
  {6175--6180} (\bibinfo {year} {2018})},\ \Eprint
  {http://arxiv.org/abs/1804.01101} {arXiv:1804.01101} \BibitemShut {NoStop}%
\bibitem [{\citenamefont {Hubbard}(1978)}]{HubbardWigner}%
  \BibitemOpen
  \bibfield  {author} {\bibinfo {author} {\bibfnamefont {J.}~\bibnamefont
  {Hubbard}},\ }\bibfield  {title} {\enquote {\bibinfo {title} {{Generalized
  Wigner lattices in one dimension and some applications to
  tetracyanoquinodimethane (TCNQ) salts}},}\ }\href {\doibase
  10.1103/PhysRevB.17.494} {\bibfield  {journal} {\bibinfo  {journal} {Phys.
  Rev. B}\ }\textbf {\bibinfo {volume} {17}},\ \bibinfo {pages} {494--505}
  (\bibinfo {year} {1978})}\BibitemShut {NoStop}%
\bibitem [{\citenamefont {{Rademaker}}\ \emph {et~al.}(2013)\citenamefont
  {{Rademaker}}, \citenamefont {{Pramudya}}, \citenamefont {{Zaanen}},\ and\
  \citenamefont {{Dobrosavljevi{\'c}}}}]{RademakerChargeOrder}%
  \BibitemOpen
  \bibfield  {author} {\bibinfo {author} {\bibfnamefont {Louk}\ \bibnamefont
  {{Rademaker}}}, \bibinfo {author} {\bibfnamefont {Yohanes}\ \bibnamefont
  {{Pramudya}}}, \bibinfo {author} {\bibfnamefont {Jan}\ \bibnamefont
  {{Zaanen}}}, \ and\ \bibinfo {author} {\bibfnamefont {Vladimir}\ \bibnamefont
  {{Dobrosavljevi{\'c}}}},\ }\bibfield  {title} {\enquote {\bibinfo {title}
  {{Influence of long-range interactions on charge ordering phenomena on a
  square lattice}},}\ }\href {\doibase 10.1103/PhysRevE.88.032121} {\bibfield
  {journal} {\bibinfo  {journal} {\pre}\ }\textbf {\bibinfo {volume} {88}}
  (\bibinfo {year} {2013}),\ 10.1103/PhysRevE.88.032121},\ \Eprint
  {http://arxiv.org/abs/1306.4765} {arXiv:1306.4765} \BibitemShut {NoStop}%
\bibitem [{\citenamefont {{Jerzy Kapcia}}\ \emph {et~al.}(2017)\citenamefont
  {{Jerzy Kapcia}}, \citenamefont {{Bara{\'n}ski}},\ and\ \citenamefont
  {{Ptok}}}]{KapciaChargeOrder}%
  \BibitemOpen
  \bibfield  {author} {\bibinfo {author} {\bibfnamefont {Konrad}\ \bibnamefont
  {{Jerzy Kapcia}}}, \bibinfo {author} {\bibfnamefont {Jan}\ \bibnamefont
  {{Bara{\'n}ski}}}, \ and\ \bibinfo {author} {\bibfnamefont {Andrzej}\
  \bibnamefont {{Ptok}}},\ }\bibfield  {title} {\enquote {\bibinfo {title}
  {{Diversity of charge orderings in correlated systems}},}\ }\href {\doibase
  10.1103/PhysRevE.96.042104} {\bibfield  {journal} {\bibinfo  {journal} {Phys.
  Rev. E}\ }\textbf {\bibinfo {volume} {96}},\ \bibinfo {pages} {042104}
  (\bibinfo {year} {2017})},\ \Eprint {http://arxiv.org/abs/1710.01757}
  {arXiv:1710.01757} \BibitemShut {NoStop}%
\bibitem [{\citenamefont {{Chen}}\ \emph {et~al.}(2004)\citenamefont {{Chen}},
  \citenamefont {{Vafek}}, \citenamefont {{Yazdani}},\ and\ \citenamefont
  {{Zhang}}}]{PairDensityWave}%
  \BibitemOpen
  \bibfield  {author} {\bibinfo {author} {\bibfnamefont {Han-Dong}\
  \bibnamefont {{Chen}}}, \bibinfo {author} {\bibfnamefont {Oskar}\
  \bibnamefont {{Vafek}}}, \bibinfo {author} {\bibfnamefont {Ali}\ \bibnamefont
  {{Yazdani}}}, \ and\ \bibinfo {author} {\bibfnamefont {Shou-Cheng}\
  \bibnamefont {{Zhang}}},\ }\bibfield  {title} {\enquote {\bibinfo {title}
  {{Pair Density Wave in the Pseudogap State of High Temperature
  Superconductors}},}\ }\href {\doibase 10.1103/PhysRevLett.93.187002}
  {\bibfield  {journal} {\bibinfo  {journal} {\prl}\ }\textbf {\bibinfo
  {volume} {93}},\ \bibinfo {eid} {187002} (\bibinfo {year} {2004})},\ \Eprint
  {http://arxiv.org/abs/cond-mat/0402323} {arXiv:cond-mat/0402323} \BibitemShut
  {NoStop}%
\bibitem [{Cfo()}]{Cfoot:renormalize}%
  \BibitemOpen
  \href@noop {} {}\bibinfo {note} {See e.g. Appendix B.1 of
  \refcite{GurarieRadzihovskyFeshbach}.}\BibitemShut {Stop}%
\bibitem [{\citenamefont {{Zhang}}\ \emph {et~al.}(2020)\citenamefont
  {{Zhang}}, \citenamefont {{Liu}},\ and\ \citenamefont
  {{Fu}}}]{FuHomobilayer}%
  \BibitemOpen
  \bibfield  {author} {\bibinfo {author} {\bibfnamefont {Yang}\ \bibnamefont
  {{Zhang}}}, \bibinfo {author} {\bibfnamefont {Tongtong}\ \bibnamefont
  {{Liu}}}, \ and\ \bibinfo {author} {\bibfnamefont {Liang}\ \bibnamefont
  {{Fu}}},\ }\bibfield  {title} {\enquote {\bibinfo {title} {{Electrically
  tunable charge transfer and charge orders in twisted transition metal
  dichalcogenide bilayers}},}\ }\href@noop {} {\  (\bibinfo {year} {2020})},\
  \Eprint {http://arxiv.org/abs/2009.14224} {arXiv:2009.14224} \BibitemShut
  {NoStop}%
\bibitem [{\citenamefont {{Angeli}}\ and\ \citenamefont
  {{MacDonald}}(2020)}]{MacDonaldHomobilayer}%
  \BibitemOpen
  \bibfield  {author} {\bibinfo {author} {\bibfnamefont {M.}~\bibnamefont
  {{Angeli}}}\ and\ \bibinfo {author} {\bibfnamefont {A.~H.}\ \bibnamefont
  {{MacDonald}}},\ }\bibfield  {title} {\enquote {\bibinfo {title}
  {{$\Gamma$-Valley Transition-Metal-Dichalcogenide Moir{\`e} Bands}},}\
  }\href@noop {} {\  (\bibinfo {year} {2020})},\ \Eprint
  {http://arxiv.org/abs/2008.01735} {arXiv:2008.01735} \BibitemShut {NoStop}%
\bibitem [{\citenamefont {{Li}}\ \emph {et~al.}(2020)\citenamefont {{Li}},
  \citenamefont {{Li}}, \citenamefont {{Naik}}, \citenamefont {{Xie}},
  \citenamefont {{Li}}, \citenamefont {{Wang}}, \citenamefont {{Regan}},
  \citenamefont {{Wang}}, \citenamefont {{Zhao}}, \citenamefont {{Zhao}},
  \citenamefont {{Kahn}}, \citenamefont {{Yumigeta}}, \citenamefont {{Blei}},
  \citenamefont {{Taniguchi}}, \citenamefont {{Watanabe}}, \citenamefont
  {{Tongay}}, \citenamefont {{Zettl}}, \citenamefont {{Louie}}, \citenamefont
  {{Wang}},\ and\ \citenamefont {{Crommie}}}]{CrommieImaging}%
  \BibitemOpen
  \bibfield  {author} {\bibinfo {author} {\bibfnamefont {Hongyuan}\
  \bibnamefont {{Li}}}, \bibinfo {author} {\bibfnamefont {Shaowei}\
  \bibnamefont {{Li}}}, \bibinfo {author} {\bibfnamefont {Mit~H.}\ \bibnamefont
  {{Naik}}}, \bibinfo {author} {\bibfnamefont {Jingxu}\ \bibnamefont {{Xie}}},
  \bibinfo {author} {\bibfnamefont {Xinyu}\ \bibnamefont {{Li}}}, \bibinfo
  {author} {\bibfnamefont {Jiayin}\ \bibnamefont {{Wang}}}, \bibinfo {author}
  {\bibfnamefont {Emma}\ \bibnamefont {{Regan}}}, \bibinfo {author}
  {\bibfnamefont {Danqing}\ \bibnamefont {{Wang}}}, \bibinfo {author}
  {\bibfnamefont {Wenyu}\ \bibnamefont {{Zhao}}}, \bibinfo {author}
  {\bibfnamefont {Sihan}\ \bibnamefont {{Zhao}}}, \bibinfo {author}
  {\bibfnamefont {Salman}\ \bibnamefont {{Kahn}}}, \bibinfo {author}
  {\bibfnamefont {Kentaro}\ \bibnamefont {{Yumigeta}}}, \bibinfo {author}
  {\bibfnamefont {Mark}\ \bibnamefont {{Blei}}}, \bibinfo {author}
  {\bibfnamefont {Takashi}\ \bibnamefont {{Taniguchi}}}, \bibinfo {author}
  {\bibfnamefont {Kenji}\ \bibnamefont {{Watanabe}}}, \bibinfo {author}
  {\bibfnamefont {Sefaattin}\ \bibnamefont {{Tongay}}}, \bibinfo {author}
  {\bibfnamefont {Alex}\ \bibnamefont {{Zettl}}}, \bibinfo {author}
  {\bibfnamefont {Steven~G.}\ \bibnamefont {{Louie}}}, \bibinfo {author}
  {\bibfnamefont {Feng}\ \bibnamefont {{Wang}}}, \ and\ \bibinfo {author}
  {\bibfnamefont {Michael~F.}\ \bibnamefont {{Crommie}}},\ }\bibfield  {title}
  {\enquote {\bibinfo {title} {{Imaging moir{\'e} flat bands in 3D
  reconstructed WSe2/WS2 superlattices}},}\ }\href@noop {} {\  (\bibinfo {year}
  {2020})},\ \Eprint {http://arxiv.org/abs/2007.06113} {arXiv:2007.06113}
  \BibitemShut {NoStop}%
\bibitem [{\citenamefont {{Rademaker}}\ and\ \citenamefont
  {{Mellado}}(2018)}]{RademakerTBG}%
  \BibitemOpen
  \bibfield  {author} {\bibinfo {author} {\bibfnamefont {Louk}\ \bibnamefont
  {{Rademaker}}}\ and\ \bibinfo {author} {\bibfnamefont {Paula}\ \bibnamefont
  {{Mellado}}},\ }\bibfield  {title} {\enquote {\bibinfo {title}
  {{Charge-transfer insulation in twisted bilayer graphene}},}\ }\href
  {\doibase 10.1103/PhysRevB.98.235158} {\bibfield  {journal} {\bibinfo
  {journal} {\prb}\ }\textbf {\bibinfo {volume} {98}},\ \bibinfo {eid} {235158}
  (\bibinfo {year} {2018})},\ \Eprint {http://arxiv.org/abs/1805.05294}
  {arXiv:1805.05294} \BibitemShut {NoStop}%
\bibitem [{\citenamefont {{Guinea}}\ and\ \citenamefont
  {{Walet}}(2018)}]{GuineaTGB}%
  \BibitemOpen
  \bibfield  {author} {\bibinfo {author} {\bibfnamefont {Francisco}\
  \bibnamefont {{Guinea}}}\ and\ \bibinfo {author} {\bibfnamefont {Niels~R.}\
  \bibnamefont {{Walet}}},\ }\bibfield  {title} {\enquote {\bibinfo {title}
  {{Electrostatic effects, band distortions, and superconductivity in twisted
  graphene bilayers}},}\ }\href {\doibase 10.1073/pnas.1810947115} {\bibfield
  {journal} {\bibinfo  {journal} {Proceedings of the National Academy of
  Science}\ }\textbf {\bibinfo {volume} {115}},\ \bibinfo {pages}
  {13174--13179} (\bibinfo {year} {2018})},\ \Eprint
  {http://arxiv.org/abs/1806.05990} {arXiv:1806.05990} \BibitemShut {NoStop}%
\bibitem [{\citenamefont {{Liu}}\ \emph {et~al.}(2019)\citenamefont {{Liu}},
  \citenamefont {{Hao}}, \citenamefont {{Khalaf}}, \citenamefont {{Lee}},
  \citenamefont {{Watanabe}}, \citenamefont {{Taniguchi}}, \citenamefont
  {{Vishwanath}},\ and\ \citenamefont {{Kim}}}]{KimPolarizedSC}%
  \BibitemOpen
  \bibfield  {author} {\bibinfo {author} {\bibfnamefont {Xiaomeng}\
  \bibnamefont {{Liu}}}, \bibinfo {author} {\bibfnamefont {Zeyu}\ \bibnamefont
  {{Hao}}}, \bibinfo {author} {\bibfnamefont {Eslam}\ \bibnamefont {{Khalaf}}},
  \bibinfo {author} {\bibfnamefont {Jong~Yeon}\ \bibnamefont {{Lee}}}, \bibinfo
  {author} {\bibfnamefont {Kenji}\ \bibnamefont {{Watanabe}}}, \bibinfo
  {author} {\bibfnamefont {Takashi}\ \bibnamefont {{Taniguchi}}}, \bibinfo
  {author} {\bibfnamefont {Ashvin}\ \bibnamefont {{Vishwanath}}}, \ and\
  \bibinfo {author} {\bibfnamefont {Philip}\ \bibnamefont {{Kim}}},\ }\bibfield
   {title} {\enquote {\bibinfo {title} {{Spin-polarized Correlated Insulator
  and Superconductor in Twisted Double Bilayer Graphene}},}\ }\href@noop {} {\
  (\bibinfo {year} {2019})},\ \Eprint {http://arxiv.org/abs/1903.08130}
  {arXiv:1903.08130} \BibitemShut {NoStop}%
\bibitem [{\citenamefont {{Chen}}\ \emph
  {et~al.}(2019{\natexlab{b}})\citenamefont {{Chen}}, \citenamefont {{Sharpe}},
  \citenamefont {{Gallagher}}, \citenamefont {{Rosen}}, \citenamefont {{Fox}},
  \citenamefont {{Jiang}}, \citenamefont {{Lyu}}, \citenamefont {{Li}},
  \citenamefont {{Watanabe}}, \citenamefont {{Taniguchi}}, \citenamefont
  {{Jung}}, \citenamefont {{Shi}}, \citenamefont {{Goldhaber-Gordon}},
  \citenamefont {{Zhang}},\ and\ \citenamefont {{Wang}}}]{WangPolarizedSC}%
  \BibitemOpen
  \bibfield  {author} {\bibinfo {author} {\bibfnamefont {Guorui}\ \bibnamefont
  {{Chen}}}, \bibinfo {author} {\bibfnamefont {Aaron~L.}\ \bibnamefont
  {{Sharpe}}}, \bibinfo {author} {\bibfnamefont {Patrick}\ \bibnamefont
  {{Gallagher}}}, \bibinfo {author} {\bibfnamefont {Ilan~T.}\ \bibnamefont
  {{Rosen}}}, \bibinfo {author} {\bibfnamefont {Eli}\ \bibnamefont {{Fox}}},
  \bibinfo {author} {\bibfnamefont {Lili}\ \bibnamefont {{Jiang}}}, \bibinfo
  {author} {\bibfnamefont {Bosai}\ \bibnamefont {{Lyu}}}, \bibinfo {author}
  {\bibfnamefont {Hongyuan}\ \bibnamefont {{Li}}}, \bibinfo {author}
  {\bibfnamefont {Kenji}\ \bibnamefont {{Watanabe}}}, \bibinfo {author}
  {\bibfnamefont {Takashi}\ \bibnamefont {{Taniguchi}}}, \bibinfo {author}
  {\bibfnamefont {Jeil}\ \bibnamefont {{Jung}}}, \bibinfo {author}
  {\bibfnamefont {Zhiwen}\ \bibnamefont {{Shi}}}, \bibinfo {author}
  {\bibfnamefont {David}\ \bibnamefont {{Goldhaber-Gordon}}}, \bibinfo {author}
  {\bibfnamefont {Yuanbo}\ \bibnamefont {{Zhang}}}, \ and\ \bibinfo {author}
  {\bibfnamefont {Feng}\ \bibnamefont {{Wang}}},\ }\bibfield  {title} {\enquote
  {\bibinfo {title} {Signatures of tunable superconductivity in a trilayer
  graphene moiré superlattice},}\ }\href {\doibase 10.1038/s41586-019-1393-y}
  {\bibfield  {journal} {\bibinfo  {journal} {Nature}\ }\textbf {\bibinfo
  {volume} {572}},\ \bibinfo {pages} {215–219} (\bibinfo {year}
  {2019}{\natexlab{b}})},\ \Eprint {http://arxiv.org/abs/1901.04621}
  {arXiv:1901.04621} \BibitemShut {NoStop}%
\bibitem [{\citenamefont {{Luo}}(2000)}]{tetron}%
  \BibitemOpen
  \bibfield  {author} {\bibinfo {author} {\bibfnamefont {Nie}\ \bibnamefont
  {{Luo}}},\ }\bibfield  {title} {\enquote {\bibinfo {title} {{Variational
  bounds on the ground-state energy of three electrons and one hole in
  two-dimension}},}\ }\href@noop {} {\  (\bibinfo {year} {2000})},\ \Eprint
  {http://arxiv.org/abs/cond-mat/0101004} {arXiv:cond-mat/0101004} \BibitemShut
  {NoStop}%
\bibitem [{Note1()}]{Note1}%
  \BibitemOpen
  \bibinfo {note} {A single parallel conductor results in \protect \mbox {$V(r)
  \propto \protect \frac {1}{r} - 1/\protect \sqrt {r^2 +
  (2d)^2}$}.}\BibitemShut {Stop}%
\bibitem [{Note2()}]{Note2}%
  \BibitemOpen
  \bibinfo {note} {The perturbative approximation $\protect \tilde {g} \sim t_1
  t_2/V_2$ formally also requires assuming $V_4 \ll V_2$ so that $V_4$ only
  results in a perturbative correction to the final state energy in
  Fig.\protect \tmspace +\thinmuskip {.1667em}\ref
  {fig:resonance}.}\BibitemShut {Stop}%
\bibitem [{Note3()}]{Note3}%
  \BibitemOpen
  \bibinfo {note} {The $\protect \sqrt {W\mu }$ prefactor in $\Delta _b$ [in
  Eq.\protect \tmspace +\thinmuskip {.1667em}(\ref {eq:Delta weak})] comes from
  the limits of integration ($-\mu $ to $W$) in Eq.\protect \tmspace
  +\thinmuskip {.1667em}(\ref {eq:DoS}). Eq.\protect \tmspace +\thinmuskip
  {.1667em}(\ref {eq:Delta weak}) is only valid when $\mu >0$. If $\mu =0$,
  then note that the ground state energy in the $\mu =0$ limit is equivalent to
  the energy in the $\mu =W$ limit if the mass is halved; i.e. $E_\protect
  \text {MF}|_{\mu =0} = E_\protect \text {MF}|_{\mu =W}^{m\to m/2}$ in
  Eq.\protect \tmspace +\thinmuskip {.1667em}(\ref {eq:MF E}). Therefore if
  $\mu =0$, then $\Delta _b \approx 2W \protect \qopname \relax o{exp}\protect
  \tmspace -\thinmuskip {.1667em}\left (-\protect \frac {\epsilon _0 - 2\mu
  }{\pi m \protect \tilde {g}^2}\right )$ [by replacing $\mu \to W$ and $m \to
  m/2$ in Eq.\protect \tmspace +\thinmuskip {.1667em}(\ref {eq:Delta weak})],
  which is significantly smaller than the expression in Eq.\protect \tmspace
  +\thinmuskip {.1667em}(\ref {eq:Delta weak}) when $0 < \mu $ and $\protect
  \frac {\epsilon _0 - 2\mu }{\pi m \protect \tilde {g}^2} \gg 1$ due to the
  missing factor of $\protect \frac {1}{2}$ in the exponent.}\BibitemShut
  {Stop}%
\bibitem [{Note4()}]{Note4}%
  \BibitemOpen
  \bibinfo {note} {A similar equation for the boson density in three spatial
  dimensions appears in Eq.\protect \tmspace +\thinmuskip {.1667em}(6.8) of
  Ref.\protect \tmspace +\thinmuskip {.1667em}\protect \rev@citealpnum
  {GurarieRadzihovskyFeshbach}.}\BibitemShut {Stop}%
\bibitem [{\citenamefont {Raikh}\ \emph {et~al.}(1996)\citenamefont {Raikh},
  \citenamefont {Glazman},\ and\ \citenamefont {Zhukov}}]{4cluster}%
  \BibitemOpen
  \bibfield  {author} {\bibinfo {author} {\bibfnamefont {M.~E.}\ \bibnamefont
  {Raikh}}, \bibinfo {author} {\bibfnamefont {L.~I.}\ \bibnamefont {Glazman}},
  \ and\ \bibinfo {author} {\bibfnamefont {L.~E.}\ \bibnamefont {Zhukov}},\
  }\bibfield  {title} {\enquote {\bibinfo {title} {{Two-Electron State in a
  Disordered 2D Island: Pairing Caused by the Coulomb Repulsion}},}\ }\href
  {\doibase 10.1103/PhysRevLett.77.1354} {\bibfield  {journal} {\bibinfo
  {journal} {Phys. Rev. Lett.}\ }\textbf {\bibinfo {volume} {77}},\ \bibinfo
  {pages} {1354--1357} (\bibinfo {year} {1996})}\BibitemShut {NoStop}%
\bibitem [{\citenamefont {Varma}(1988)}]{VarmaMissingValence}%
  \BibitemOpen
  \bibfield  {author} {\bibinfo {author} {\bibfnamefont {C.~M.}\ \bibnamefont
  {Varma}},\ }\bibfield  {title} {\enquote {\bibinfo {title} {{Missing valence
  states, diamagnetic insulators, and superconductors}},}\ }\href {\doibase
  10.1103/PhysRevLett.61.2713} {\bibfield  {journal} {\bibinfo  {journal}
  {Phys. Rev. Lett.}\ }\textbf {\bibinfo {volume} {61}},\ \bibinfo {pages}
  {2713--2716} (\bibinfo {year} {1988})}\BibitemShut {NoStop}%
\end{thebibliography}%

\appendix


\newpage

\section{Additional Details}
\label{app:details}

\begin{figure}
  \includegraphics[width=.6\columnwidth]{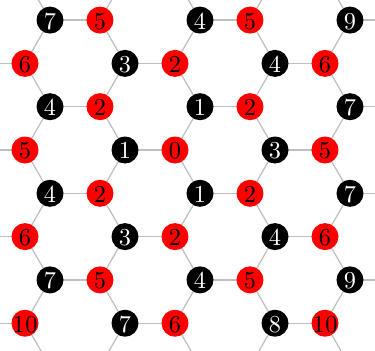}
  \caption{%
    $n^\text{th}$ nearest-neighbors on a honeycomb lattice.
    Throughout this work, $V_n$ denotes the Coulomb repulsion between $n^\text{th}$ nearest-neighbors.
    We sometimes differentiate between $V_n^{AA}$ and $V_n^{BB}$ for repulsions between a pair of sites on the $A$ or $B$ sublattice.
    $n^\text{th}$ nearest-neighbors are separated by $a_n$ with $a_{n=1,2,3,4,5}/a_1 = 1, \sqrt{3}, 2, \sqrt{7}, 3$.
    $L_\text{M} = a_2$ is the moir\'e period.
    }\label{fig:neighbors}
\end{figure}

In this appendix, we will provide extra details on the low-energy excitations,
  followed by a mean-field theory analysis of the superconducting states starting from the lattice model in \eqnref{eq:H}.
We will consider the low density and strongly interacting limit:
\begin{equation}
  n-1 = \delta \ll 1 \quad\text{and}\quad t_{ij} \ll V_2
\end{equation}
where $V_n$ is the Coulomb repulsion between $n$-th nearest-neighbors (see \figref{fig:neighbors}).
We also assume that the spins are fully polarized, e.g. by an external magnetic field.

\subsection{Electrostatics}
\label{app:electrostatics}

\begin{figure}
  \subfloat[]{\includegraphics[width=.79\columnwidth]{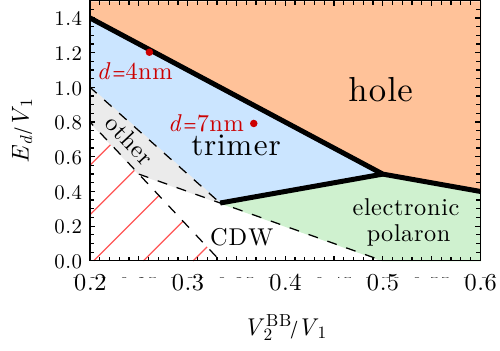}} \hspace{.02\columnwidth}
  \subfloat[]{\raisebox{2cm}{\includegraphics[width=.17\columnwidth]{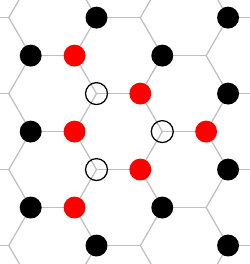}}}
  \caption{%
    (a) A detailed excitation phase diagram of the lowest energy per charge ($E/q$) excitation.
    The red dots mark estimates for a slightly twisted WSe$_2$/WS$_2$ inside a dielectric environment with permittivity $\epsilon=3$ and for different distances $d$ to metallic gates.
    In the ``other'' region, other excitations have the least $E/q$, such as the charge-$3e$ excitation shown in (b).
    The bottom-left region (white with red stripes) is inaccessible (when $V_{n\geq3}=0$) as it would require $\Delta<0$.
    The dashed lines are drawn for $V_{n\geq3}=0$.
    See \appref{app:electrostatics} for more details.
  }\label{fig:detailedExcitations}
\end{figure}

\begin{figure}
  \subfloat[\label{fig:dipoleCrystal}]{\includegraphics[width=.45\columnwidth]{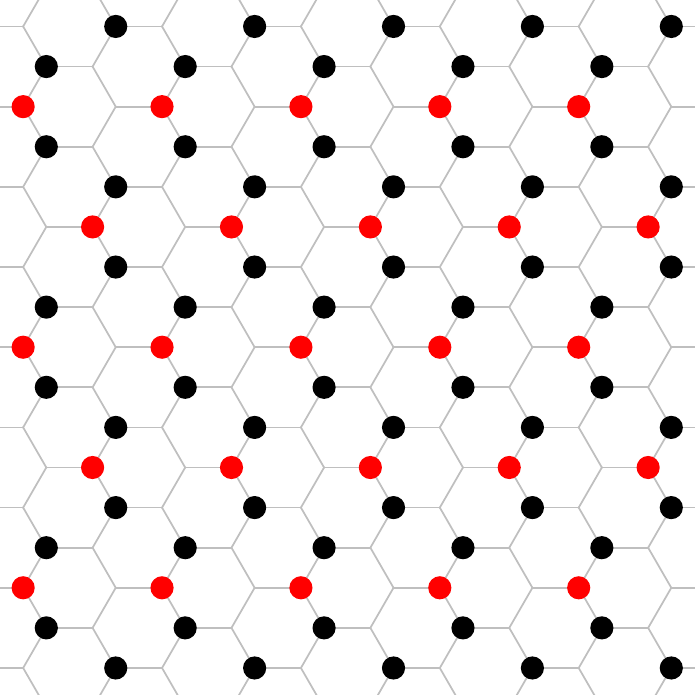}} \hspace{.05\columnwidth}
  \subfloat[\label{fig:snake}]{\includegraphics[width=.45\columnwidth]{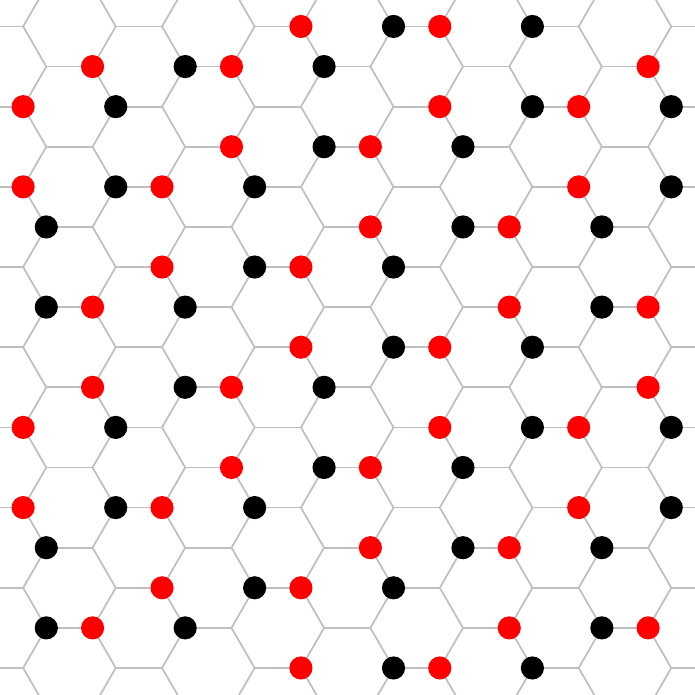}}
  \caption{%
    Possible instabilities of the charge transfer insulator.
    (a) A Wigner crystal of dipoles, which is a possible state when the dipole energy is negative, $E_d<0$.
    (b) The charge density wave that occurs in the CDW region (white) of \figref{fig:detailedExcitations} when $V_{n\geq4}=0$ and $V_3 \ll V_2$.
    }\label{fig:CDW1}
\end{figure}

In \figref{fig:excitation} of the main text, we showed which of the following three excitations has the smallest energy per charge, $E/q$:
\begin{align}
  \text{hole ($q=1$):} & & E_{-e} & \label{eq:Eq}\\
  \text{trimer ($q=2$):} & & E_t  &= 2E_e + E_d - 2V_1 + 3V_2^{BB} \nonumber\\
  \begin{array}{c} \text{electronic} \\ \text{polaron} \end{array} (q=1): & & E_{ep} &= E_e + E_d - V_1 + V_2^{BB} \nonumber
\end{align}
Remarkably, for the above three excitations, this only requires knowing $E_d/V_1$ and $V_2^{BB}/V_1$;
  $\Delta$ and all $V_n$ are effectively absorbed into these two ratios.

It is remarkable that charge-$2e$ pairing can occur from just repulsive interactions.
The nontrivial charge-transfer insulating background is essential for the trimer stability;
  a trimer is not stable in the vacuum \cite{tetron}.
As an aid to intuition,
  we give an even simpler example of how this can occur in a finite system in \appref{app:cluster}.

In \figref{fig:detailedExcitations}, we show a more detailed diagram of the smallest $E/q$ excitation when arbitrary excitations are considered.
We also check for instabilities (shown in \figref{fig:CDW1}) of the charge transfer insulator,
  which occur in the ``CDW'' region of \figref{fig:detailedExcitations} and when $E_d<0$.
The energy of other excitations and these instabilities depend on more than just $E_d/V_1$ and $V_2^{BB}/V_1$.
Therefore, we show the locations of these other excitations and instabilities for the simple case when $V_{n\geq3}=0$.
In \figref{fig:detailedExcitations}, dashed lines are used to depict boundaries that assume $V_{n\geq3}=0$.

\begin{table}
  \subfloat[$d = 4\text{nm}$]{\begin{tabular}{c|cccccc}
    $n$      & 0      & 1      & 2      & 3      \\\hline
  $V_n^{AA}$ & 3.7769 &        & 0.2292 &        \\
  $V_n^{AB}$ &        & 0.9479 &        & 0.1340 \\
  $V_n^{BB}$ & 2.9828 &        & 0.2472 &
  \end{tabular}} \\
  \subfloat[$d = 7\text{nm}$]{\begin{tabular}{c|cccccc}
    $n$      & 0      & 1      & 2      & 3      \\\hline 
  $V_n^{AA}$ & 4.2407 &        & 0.4599 &        \\ 
  $V_n^{AB}$ &        & 1.2998 &        & 0.3239 \\ 
  $V_n^{BB}$ & 3.4132 &        & 0.4780 &           
  \end{tabular}} \\
  \caption{%
    The values of $V_n$ (in units of $\frac{e^2}{\epsilon L_\text{M}} = \frac{205.7}{\epsilon}\text{meV}$ with $L_\text{M} = 7\text{nm}$) used to estimate the location (denoted by red dots in \figref{fig:detailedExcitations})
      of slightly twisted WSe$_2$/WS$_2$ in the excitation phase diagram.
  }\label{tab:Vn}
\end{table}

We also estimate where in the phase diagram a slightly twisted WSe$_2$/WS$_2$ with a moir\'e period $L_\text{M} = 7\text{nm}$ could be,
  which we show in \figref{fig:detailedExcitations} using red dots.
The locations are calculated using $\Delta = 14.9\text{meV}$ and the values of $V_n$ shown in \tabref{tab:Vn}.
$\Delta$ and $V_n$ were calculated using Wannier orbitals
  and a Coulomb interaction $V(r)$ that is screened by a pair of metallic gates,
  each a distance $\pm d$ from the TMD hereobilayer. \cite{LiuZhandPrivate}
By modeling the gates as perfect conductors, the screened Coulomb interaction can be calculated using the method of image charges, yielding\footnote{%
  A single parallel conductor results in \mbox{$V(r) \propto \frac{1}{r} - 1/\sqrt{r^2 + (2d)^2}$}.}
\begin{equation}
  V(r) = \frac{e^2}{\epsilon} \sum_{z \in \mathbb{Z}} \, \frac{(-1)^z}{\sqrt{r^2 + (2dz)^2}} \label{eq:2gate}
\end{equation}
When $r \gg d$, $V(r)$ decays exponentially.

\begin{figure}
  \includegraphics[width=\columnwidth]{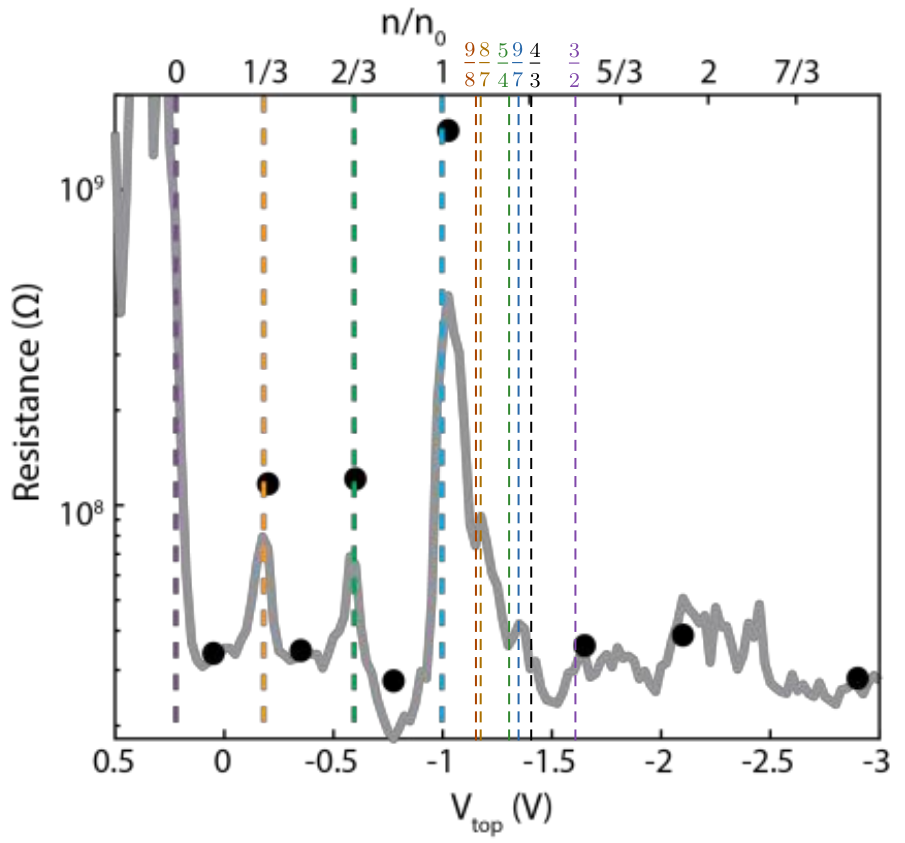} 
  \caption{%
    Resistance of the Berkeley group's WSe$_2$/WS$_2$ moir\'e device as a function of gate voltage,
      which determines the electron filling, $n$.
    The figure is copied from \refcite{BerkeleyTMD}.
    We add vertical lines to indicate various Wigner crystal fillings.
    The blue line at filling $n = 9/7$ is particularly interesting because it shows a large resistance peak at the same filling as the pair density wave in \figref{fig:9_7_pairCrystal}.
    Energetically favorable charge density waves at filling $n=3/2$ are shown in \figref{fig:CDW 1_2}.
    }\label{fig:BerkeleyData}
\end{figure}

In \figref{fig:BerkeleyData}, we point out possible experimental evidence of insulating pair density waves of trimers from recently observed resistivity peaks \cite{BerkeleyTMD}.
See also \refcite{ShanTMD} for evidence for additional charge orders.

\begin{figure}
  \subfloat[$\delta = 1/2$]{\includegraphics[width=.45\columnwidth]{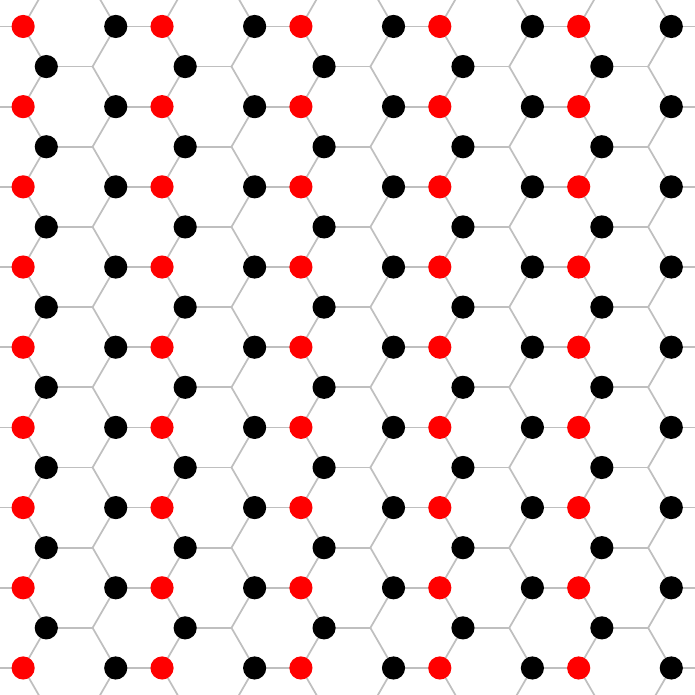}} \hspace{.05\columnwidth}
  \subfloat[$\delta = 1/2$]{\includegraphics[width=.45\columnwidth]{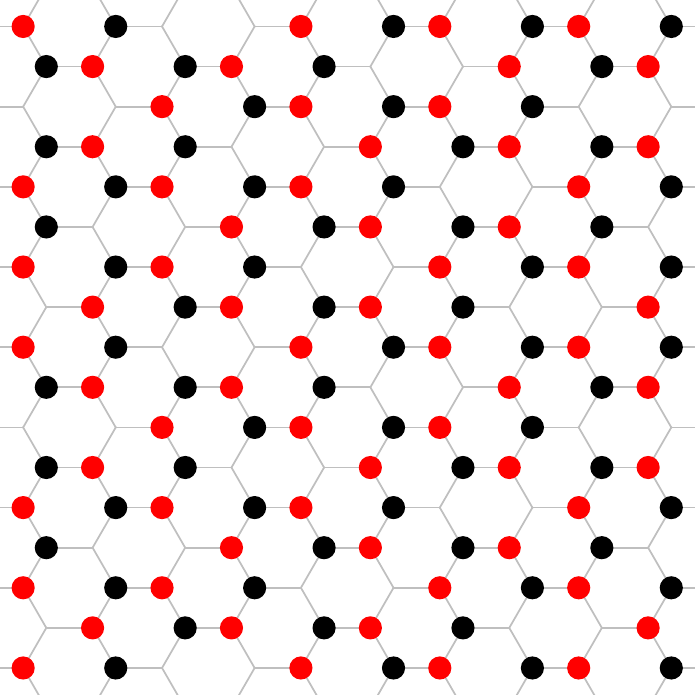}} \\
  \caption{%
    A charge density wave state at doping $\delta=1/2$ that is typical of the (a) hole and (b) trimer regimes of the excitation phase diagram in \figref{fig:detailedExcitations}.
  }\label{fig:CDW 1_2}
\end{figure}

\subsection{Mean Field Theory of Superconductivity}
\label{app:MFT}

Here, we study the mean field theory of trimer superconductivity.
Suppose that we are near the edge of the trimer region of the phase diagram,
  so that the trimer binding energy, $\epsilon_b$ [\eqnref{eq:Eb}], is small:
\begin{equation}
  0 < \epsilon_b \ll V_2
\end{equation}
We will also assume that excitations other than the charge-$e$ hole and charge-$2e$ trimers (such as the dipole and electronic polaron) have a large energy cost $\sim V_2$ so that they can be ignored.

The low-energy Hamiltonian thus consists of only the mobile holes $c_{\bf k}$ on the cations,
  and bosonic trimers $b_a$ centered on the anions:
\begin{align}
H_\text{eff} &= \sum_{\bf k} \epsilon({\bf k}) c^\dagger_{\bf k} c_{\bf k} + (\epsilon_0 - 2\mu) \sum_a b_a^\dagger b_a \nonumber\\
             &- \frac{\tilde{g}({\bf k-k'})}{2\sqrt{N}} \sum_{\bf k,k'} (b_{\bf k+k'}^\dagger c_{\bf k} c_{\bf k'} + h.c.) + \cdots \label{eq:HeffApp}
\end{align}
where $b_{\bf k}^\dagger = N^{-1/2} \sum_a e^{i {\bf k}\cdot a} b_a^\dagger$ and $N$ is the number of anion sites.

The first term gives the dispersion of the holes, which hop on the triangular lattice of cations.
At low density~($\delta \ll 1$), $\epsilon({\bf k})$ can be expanded about its band minima $\pm {\bf K} = (0,\pm\frac{4\pi}{3 L_\text{M}})$:
\begin{equation}
  \epsilon({\bf k \pm K}) = \frac{1}{2m} k^2 - \mu + O(k^3) \label{eq:epsilon}
\end{equation}
$m \approx 2L_\text{M}^2/27t_2$ where $t_n$ is the hopping energy between $n$-th nearest-neighbors.
We have shifted $\epsilon({\bf k})$ so that $\epsilon(\pm {\bf K}) = 0$.

The second term sets the energy cost of trimers.
$b^\dagger_a$ creates a trimer centered at the anion $a$,
  and $b^\dagger_{\bf k}$ creates a trimer with momentum $\bf k$.
$\epsilon_0 \approx 6t_2 - \epsilon_b$ is the energy difference between a trimer and two holes at the band minima $\pm {\bf K}$.

The last term accounts for the conversion between a trimer and a pair of holes.
The triangular symmetry of the cation sublattice constrains the conversion amplitude to the following form:
  $\tilde{g}({\bf q}) = \tilde{g} \sum_{i=1}^3 \frac{2}{3\sqrt{3}} \sin({\bf q} \cdot {\bf r}_i) + \cdots$
  where ``$\cdots$'' denotes higher-order moments and
  ${\bf r}_i = \sqrt{3} \sin(\frac{2\pi}{3} i) \hat{\bf x} - \sqrt{3} \cos(\frac{2\pi}{3} i) \hat{\bf y}$
  are second-nearest-neighbor displacements.
The normalization is such that $\tilde{g}({\bf K}) = - \tilde{g}(- {\bf K}) = \tilde{g}(-2{\bf K}) = \tilde{g}$.
$\tilde{g}$ can be calculated perturbatively;
  the leading contribution comes from the process shown in \figref{fig:resonance} and has amplitude $\tilde{g} \sim t_1 t_2/V_2$.\footnote{%
  The perturbative approximation $\tilde{g} \sim t_1 t_2/V_2$ formally also requires assuming $V_4 \ll V_2$ so that $V_4$ only results in a perturbative correction to the final state energy in \figref{fig:resonance}.}

Expanding $H_\text{eff}$ about the band minima $\bf \pm K$ leads to
\begin{align}
H_\text{eff} &\approx \sum_{\bf k, \pm} \epsilon_k c^\dagger_{\bf k, \pm} c_{\bf k, \pm} + (\epsilon_0 - 2\mu) \sum_a b_a^\dagger b_a \nonumber\\
             &- \frac{\tilde{g}}{\sqrt{N}} \sum_{\bf k,k'} (b_{\bf k+k'}^\dagger c_{\bf k, -} c_{\bf k', +} + h.c.) + \cdots \label{eq:HeffApp2}
\end{align}
where $c_{{\bf k},\pm} \sim c_{\bf k \pm K}$ denotes the new fermion operators expanded about $\bf \pm K$
  and $\epsilon_k = \frac{1}{2m} k^2 - \mu$ is the dispersion.

The ``$\cdots$'' in $H_\text{eff}$ denotes other terms that could be included in $H_\text{eff}$.
We will ignore these terms in the following mean-field analysis because we do not expect these terms to be relevant in the resonantly-paired superconductivity regime of interest.
To justify this, consider two potentially important kinds terms that we are omitting.
The first is a trimer kinetic energy term, $-t_t \sum_{a'a} b_{a'}^\dagger b_a$,
  where $t_t \sim t^3/V^2$ is the trimer hopping energy, resulting from the perturbative process shown in \figref{fig:BEC}.
However, we expect that near resonance, this term is negligible compared to the effective boson mass resulting from the coupling $\tilde{g}$ to the fermions.
The second potentially important terms are 4-fermion interactions, such as $V_{ij} n_i n_j$.
However, we will soon see [from \eqnref{eq:Delta resonant}] that near resonance,
  the fermion and boson operators scale as $c \sim b \sim O(\sqrt{\delta})$ at low density $\delta$.
Therefore the first term in $H_\text{eff}$ is $O(t \delta)$; the third term contributes $O(\delta^{3/2} \tilde{g}) \sim O(\delta^{3/2} t^2/V_2)$;
  and a 4-fermion interaction would contribute $O(\delta^2 V)$.
Thus, we expect that the 4-fermion interaction is negligible when $\delta^2 V \ll \delta^{3/2} t^2/V$,
  i.e. at the sufficiently-low doping $\delta \ll (t/V)^4$.

To make a connection to $H_\delta$ in \eqnref{eq:Hdelta} of the main text,
  note that $\psi^\dagger \sim L_\text{M}^{-1} c^\dagger$ and
  $\phi^\dagger \sim L_\text{M}^{-1} b^\dagger$.
Then $g$ in $H_\delta$ and $\tilde{g}$ in $H_\text{eff}$ [\eqnref{eq:HeffApp2}] are related by $g \sim L_\text{M} \tilde{g}$.
In the following, we will use lattice units where the distance between nearest-neighbor sites is $1$,
  so that the distance between next-nearest-neighbors is $L_\text{M} = \sqrt{3}$.

To make analytical progress, we consider the following mean-field approximation:
\begin{gather}
\begin{aligned}
  b_a^\dagger b_a &=       b_a^\dagger \langle b_a \rangle + \langle b_a^\dagger \rangle b_a - \langle b_a^\dagger \rangle \langle b_a \rangle + (b_a^\dagger - \langle b_a^\dagger \rangle) (b_a - \langle b_a \rangle) \nonumber\\
                  &\approx b_a^\dagger \langle b_a \rangle + \langle b_a^\dagger \rangle b_a - \langle b_a^\dagger \rangle \langle b_a \rangle
\end{aligned} \\
b_{\bf k+k'}^\dagger c_{\bf k,-} c_{\bf k',+} \approx \langle b_{\bf k+k'}^\dagger \rangle c_{\bf k,-} c_{\bf k',+}
\end{gather}
With this approximation, the low-energy Hamiltonian becomes quadratic:
\begin{gather}
\begin{aligned}
  H_\text{MF} =& \sum_{\bf k} \begin{pmatrix} c_{\bf +k,+} \\ c_{-\bf k,-}^\dagger \end{pmatrix}^\dagger
                 \begin{pmatrix} +\epsilon_k & -\Delta_b \\ -\Delta_b & -\epsilon_k \end{pmatrix}
                 \begin{pmatrix} c_{\bf +k,+} \\ c_{-\bf k,-}^\dagger \end{pmatrix} \\
              &\;\;\;+ \frac{\epsilon_0 - 2\mu}{\tilde{g}^2} \Delta_b^2
\end{aligned} \\
  \Delta_b = \tilde{g} \langle b_a \rangle \nonumber
\end{gather}
$\Delta_b$ is the superconducting order parameter.
$\Delta_b > 0$ is assumed to be positive (without loss of generality).

The ground state energy density is
\begin{align}
  \frac{E_\text{MF}}{N} &= - \int_E D(E) \sqrt{E^2 + \Delta_b^2} + \frac{\epsilon_0 - 2\mu}{\tilde{g}^2} \Delta_b^2 \\
  D(E) &= \int_{\bf k} \delta(E - \epsilon_k) = \begin{cases} 2\pi m & -\mu < E < W \\ 0 & \text{otherwise} \end{cases} \label{eq:DoS}
\end{align}
where $D(E)$ is the density of single-particle states, and
  $\int_{\bf k} = \int \frac{d^3{\bf k}}{(2\pi)^2} \Theta(W - \epsilon_k)$ integrates over momentum states with energy $\epsilon_k < W$.
$W$ is a UV cutoff which
  can be taken to be $W = (2\pi m)^{-1} - \mu \approx (2\pi m)^{-1}$ so that $\int_E D(E) = 1$; 
  this is
  roughly equal to the bandwidth $9t_2 \approx 2m^{-1}$.
Evaluating the integral yields
\begin{align}
  \frac{E_\text{MF}}{N} = -\pi m &\Bigg[ W \sqrt{W^2 + \Delta_b^2}+ \Delta_b^2 \log\!\left(\!W + \sqrt{W^2 + \Delta_b^2}\right) \nonumber\\
    &                \!\!\!+\mu\; \sqrt{\,\mu^2\, + \, \Delta_b^2} + \Delta_b^2 \log\!\left(\mu\, +\, \sqrt{\,\mu^2\, +\, \Delta_b^2}\right) \nonumber\\
    &  \!\!\!- 2\Delta_b^2 \log\Delta_b \Bigg] + \frac{\epsilon_0 - 2\mu}{\tilde{g}^2} \Delta_b^2 \label{eq:MF E}
\end{align}

The superconducting order parameter $\Delta_b$ can be calculated by minimizing the energy as a function of $\Delta_b$,
  which yields
\begin{equation}
  \Delta_b = \dfrac{\sqrt{W^2 + \mu^2 + 2W\mu \cosh\frac{\epsilon_0 - 2\mu}{\pi m \tilde{g}^2}}}{\sinh\frac{\epsilon_0 - 2\mu}{\pi m \tilde{g}^2}}
\end{equation}

$\Delta_b$ depends strongly on the chemical potential $\mu$,
  which can be obtained from the filling constraint:
\begin{align}
  \delta   &= \delta_c + 2\delta_b \nonumber \\
  \delta_c &= \langle c_i^\dagger c_i \rangle = 4\pi m \mu \label{eq:delta MF}\\
  \delta_b &= \langle b_a^\dagger b_a \rangle \approx \frac{\Delta_b^2}{\tilde{g}^2} \nonumber
\end{align}
$\delta_c$ is the density of holes.
$\delta_b$ is the density of bosonic trimers, which we approximate at the mean-field level:
  $\langle b_a^\dagger b_a \rangle \approx \langle b_a^\dagger \rangle \langle b_a \rangle = \Delta_b^2/\tilde{g}^2$.

There are two regimes:
  (1) BCS superconductivity when $\frac{\epsilon_0 - 2\mu}{\pi m \tilde{g}^2} \gg 1$, and
  (2) resonantly-paired superconductivity when $\epsilon_0 \approx 2\mu$.

\subsubsection{BCS Superconductivity Regime}

When $\frac{\epsilon_0 - 2\mu}{\pi m \tilde{g}^2} \gg 1$,
  the order parameter can be approximated as
\begin{equation}
  \Delta_b \approx 2\sqrt{W\mu} \, \exp\!\left(-\frac{\epsilon_0 - 2\mu}{2\pi m \tilde{g}^2}\right) \label{eq:Delta weak}
\end{equation}
and a BCS superconductivity regime occurs where $\Delta_b$ is very small.\footnote{%
  The $\sqrt{W\mu}$ prefactor in $\Delta_b$ [in \eqnref{eq:Delta weak}] comes from the limits of integration ($-\mu$ to $W$) in \eqnref{eq:DoS}.
  \eqnref{eq:Delta weak} is only valid when $\mu>0$.
  If $\mu=0$, then note that the ground state energy in the $\mu=0$ limit is equivalent to the energy in the $\mu=W$ limit if the mass is halved;
    i.e. $E_\text{MF}|_{\mu=0} = E_\text{MF}|_{\mu=W}^{m\to m/2}$ in \eqnref{eq:MF E}.
  Therefore if $\mu=0$, then $\Delta_b \approx 2W \exp\!\left(-\frac{\epsilon_0 - 2\mu}{\pi m \tilde{g}^2}\right)$ [by replacing $\mu\to W$ and $m \to m/2$ in \eqnref{eq:Delta weak}],
    which is significantly smaller than the expression in \eqnref{eq:Delta weak} when $0 < \mu$ and $\frac{\epsilon_0 - 2\mu}{\pi m \tilde{g}^2} \gg 1$ due to the missing factor of $\frac{1}{2}$ in the exponent.}
As a result, the boson density is very small ($\delta_b \ll \delta$),
  which allows us to approximately solve for the chemical potential from \eqnref{eq:delta MF}:
\begin{equation}
  \mu \approx \frac{\delta}{4\pi m}
\end{equation}

This regime is very similar to BCS superconductivity.
This can be understood by integrating out the boson to obtain a 4-fermion interaction
  $\tilde{g}' c_+^\dagger c_-^\dagger c_- c_+$ with $\tilde{g}' \sim \frac{\tilde{g}^2}{\epsilon_0 - 2\mu}$.
In terms of $\tilde{g}'$, the order parameter $\Delta_b$ scales exactly the same as the BCS order parameter (in two spatial dimensions):
  $\Delta_b \sim \Delta_\text{BCS} \sim \sqrt{W\mu} e^{-1/D \tilde{g}'}$,
  where $D = 2\pi m$ is the density of states from \eqnref{eq:DoS}.

Note that in this regime, the boson density is very small, so the $\tilde{g}$ coupling term in $H_\text{eff}$ [\eqnref{eq:HeffApp2}] contributes very little to the energy.
Therefore in this regime, the terms in the ``$\cdots$'' of $H_\text{eff}$ are likely to play an important role and possibly result in other kinds of symmetry breaking.
So although BCS-like superconductivity results when the ``$\cdots$'' terms are dropped,
  a more detailed analysis is needed to determine the true ground state in this regime when the ``$\cdots$'' terms are included.

\subsubsection{Resonantly-Paired Superconductivity Regime}

The boson density diverges as the chemical potential approaches its maximum value: $\mu \to \epsilon_0/2$.
Approximating $\mu \approx \epsilon_0/2$ allows us to solve for the boson density $\delta_b$ in \eqnref{eq:delta MF}, which can be used to express the order parameter:\footnote{%
  A similar equation for the boson density in three spatial dimensions appears in Eq.\,(6.8) of \refcite{GurarieRadzihovskyFeshbach}.}
\begin{align}
  \Delta_b &\approx \tilde{g} \sqrt{\delta_b} \label{eq:Delta resonant}\\
  \delta_b &\approx \frac{1}{2} \delta - \pi m \epsilon_0 \nonumber
\end{align}
Due to the significantly larger boson density,
  the order parameter $\Delta_b \sim \tilde{g}$ is immensely larger in this resonantly-paired regime than in the BCS regime where $\Delta_b \sim e^{-1/\tilde{g}^2}$ [\eqnref{eq:Delta weak}].

\newpage

\section{Valence Skipping in a 4-site Cluster}
\label{app:cluster}

A toy model for trimer stability is obtained by considering a 4-site cluster \cite{4cluster} of the Hamiltonian $H_0$ in \eqnref{eq:H} at a chemical potential $\mu$:
\begin{equation}
 \begin{split} \includegraphics{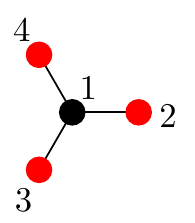} \end{split}
 \quad
 \begin{split}
 H_4 &= \Delta \, (n_2 + n_3 + n_4) \\ &+ V_1 \, n_1 (n_2 + n_3 + n_4) \\
     &+ V_2 \, (n_2 n_3 + n_3 n_4 + n_4 n_2) \\ &- \mu \, (n_1 + n_2 + n_3 + n_4)
 \end{split}
 \label{eq:H4}
\end{equation}
Each site either has 0 or 1 fermions, $n_i = 0,1$,
 which is physically relevant when a large on-site Hubbard interaction prevents double occupancy.

\begin{figure}
 \includegraphics[width=\columnwidth]{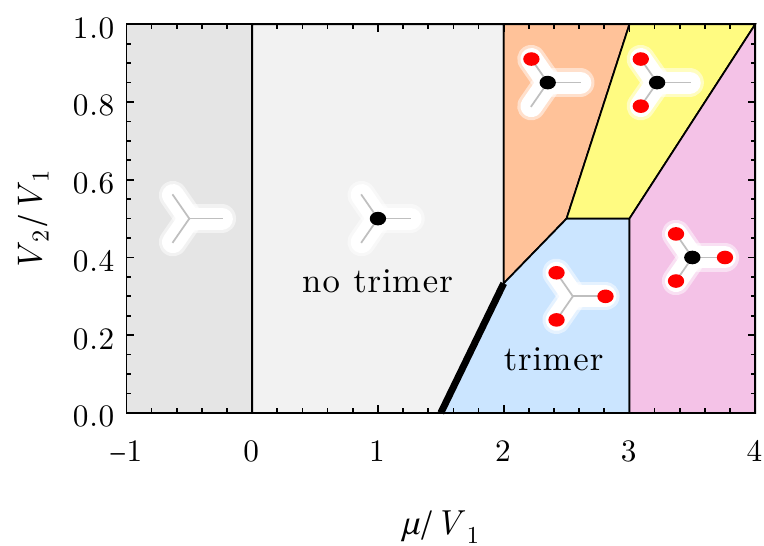}
 \caption{%
   Ground states of the 4-site cluster model in \eqnref{eq:H4} with $\Delta = V_1$.
   Valence skipping (where charge of the ground state jumps by two) \cite{VarmaMissingValence} occurs along the thick line.
   Black and red dots denote filled orbitals with $n_i=1$.
   }\label{fig:4site}
\end{figure}

The ground state phase diagram of the 4-site cluster is shown in \figref{fig:4site}.
The ``no trimer'' region is analogous to a change transfer insulator,
  while the ``trimer'' region is analogous to a trimer excitation of the change transfer insulator.

The ground states are easiest to understand in an ideal limit where $V_2=0$, $\Delta=V_1$, and $\mu=\frac{3}{2} V_1$.
Then $H_4 = V_1 \, (n_1 - \frac{1}{2})(n_2 + n_3 + n_4 - \frac{3}{2})$,
 and it is simple to see that the ``no trimer'' and ``trimer'' states are degenerate ground states.


\end{document}